\pdfoutput=1
\newcommand*{\ATLASTEMPLATEPATH}{atlaslatex/}
\newcommand*{\ATLASLATEXPATH}{\ATLASTEMPLATEPATH latex/}
\newcommand*{\ATLASLOGOSPATH}{\ATLASTEMPLATEPATH logos/}

\documentclass[cernpreprint, UKenglish, texlive=2016,titleextraheight=13mm]{\ATLASLATEXPATH atlasdoc}

\usepackage[backend=biber, block=space]{\ATLASLATEXPATH atlaspackage}
\usepackage{\ATLASLATEXPATH atlasbiblatex}

\usepackage{\ATLASLATEXPATH atlascontribute}

\usepackage{\ATLASLATEXPATH atlasphysics}

\makeatletter
\DeclareOldFontCommand{\rm}{\normalfont\rmfamily}{\mathrm}
\DeclareOldFontCommand{\sf}{\normalfont\sffamily}{\mathsf}
\DeclareOldFontCommand{\tt}{\normalfont\ttfamily}{\mathtt}
\DeclareOldFontCommand{\bf}{\normalfont\bfseries}{\mathbf}
\DeclareOldFontCommand{\it}{\normalfont\itshape}{\mathit}
\DeclareOldFontCommand{\sl}{\normalfont\slshape}{\@nomath\sl}
\DeclareOldFontCommand{\sc}{\normalfont\scshape}{\@nomath\sc}
\makeatother

\graphicspath{{\ATLASLOGOSPATH}{figures/}}

\newcommand{\myatlaspaper}{ATLAS}

\makeatletter
\newif\ifallloaded
\newcommand{\IfPackagesLoaded}[3]{  \allloadedtrue
  \@for\@tempa:=#1\do{    \ltx@ifpackageloaded{\@tempa}{}{\allloadedfalse}}  \ifallloaded #2\else #3\fi}
\makeatother

\ifx\myatlaspaper\undefined
    \newcommand{\lhcenergySeven}{\ensuremath{ 7 \TeV}}
    \newcommand{\lhcenergyEight}{\ensuremath{ 8 \TeV}}
    \newcommand{\lhcenergyComb}{\ensuremath{ 7 \text{ and } 8 \TeV}}
\else
    \newcommand{\lhcenergySeven}{\ensuremath{ 7~\TeV}}
    \newcommand{\lhcenergyEight}{\ensuremath{ 8~\TeV}}
    \newcommand{\lhcenergyComb}{\ensuremath{ 7 \text{ and } 8~\TeV}}
\fi
\newcommand{\cmenergySeven}{\ensuremath{\sqrt{s}=\lhcenergySeven}}
\newcommand{\cmenergyEight}{\ensuremath{\sqrt{s}=\lhcenergyEight}}
\newcommand{\cmenergyComb}{\ensuremath{\sqrt{s}=7 \text{ and } \lhcenergyEight}}

\newcommand{\makebold}[2][n]{  \mbox{
  \mathcode`,=\numexpr\mathcode`,-"6000
    \if#1b\boldmath\fi
    $#2$    }
}

\newcolumntype{L}[1]{D{.}{#1}{1,3}}

\newcommand{\mytt}{\ensuremath{t \bar{t}}\xspace}
\newcommand{\mytbar}{\ensuremath{\bar{t}}\xspace}
\newcommand{\mybbar}{\ensuremath{\bar{b}}\xspace}

\newcommand{\myt}{\ensuremath{t}\xspace}
\newcommand{\myb}{\ensuremath{b}\xspace}
\newcommand{\mys}{\ensuremath{s}\xspace}

\newcommand{\myc}{\ensuremath{c}\xspace}

\newcommand{\myq}{\ensuremath{q}\xspace}

\newcommand{\mytW}{\ensuremath{tW}\xspace}
\newcommand{\myWt}{\ensuremath{tW}\xspace}
\newcommand{\myW}{\ensuremath{W}\xspace}

\newcommand{\mypp}{\ensuremath{pp}\xspace}
\newcommand{\myppbar}{\ensuremath{{p}\bar{p}}\xspace}
\newcommand{\mytWb}{\ensuremath{tWb}\xspace}
\newcommand{\mytWd}{\ensuremath{tWd}\xspace}
\newcommand{\mytWs}{\ensuremath{tWs}\xspace}

\newcommand{\myWjets}{\ensuremath{W}+jets\xspace}
\newcommand{\myZjets}{\ensuremath{Z}+jets\xspace}
\newcommand{\myWZjets}{\ensuremath{W/Z}+jets\xspace}

\newcommand{\vtd}{\ensuremath{V_{td}}\xspace}
\newcommand{\vts}{\ensuremath{V_{ts}}\xspace}
\newcommand{\vtb}{\ensuremath{V_{\myt\myb}}\xspace}

\newcommand*{\hathor}{\textsc{HatHor}\xspace}

\newcommand{\absvtb}{\ensuremath{|f_{\rm LV}\vtb|}\xspace}
\newcommand{\vtbsq}{\ensuremath{\absvtb^2}\xspace}

\newcommand{\myMET}{\ensuremath{E_{\mathrm{T}}^{\mathrm{miss}}}\xspace}
\newcommand{\mypT}{\ensuremath{p_{\mathrm{T}}}\xspace}

\newcommand{\tch}{\myt-channel\xspace}
\newcommand{\sch}{\mys-channel\xspace}

\newcommand{\topmass}{m_{\myt}\xspace}

\newcommand{\invfb}{\ensuremath{\rm fb^{-1}}\xspace}

\newcommand{\flv}{\ensuremath{f_{\rm LV}}\xspace}

\newcommand{\sigmameas}{\ensuremath{\sigma_{\rm meas.}}\xspace}
\newcommand{\sigmatheo}{\ensuremath{\sigma_{\rm theo.}}\xspace}

\newcommand{\sigmatch}{\ensuremath{\sigma_{{\myt\textrm{-chan.}}}}\xspace}
\newcommand{\sigmasch}{\ensuremath{\sigma_{\mys\textrm{-chan.}}}\xspace}
\newcommand{\sigmatW}{\ensuremath{\sigma_{\mytW}}\xspace}
\newcommand{\sigmatchtheo}{\ensuremath{\sigma^{\myt\textrm{-chan.}}_{\mathrm{\textrm{theo.}}}}\xspace}
\newcommand{\sigmaschtheo}{\ensuremath{\sigma^{\mys\textrm{-chan.}}_{\mathrm{\textrm{theo.}}}}\xspace}
\newcommand{\sigmatWtheo}{\ensuremath{\sigma^{\mytW}_{\mathrm{\textrm{theo.}}}}\xspace}

\newcommand\T{\rule{0pt}{2.6ex}}              \newcommand\B{\rule[-1.2ex]{0pt}{0pt}}

 \newcommand{\finalResulttChanSeven}{\ensuremath{\sigmatch = 67.5\pm2.4\;(\text{stat.})\pm5.0\;(\text{syst.})\pm1.1\;(\text{lumi.})\,\,\text{pb} = 67.5\pm5.7\,\,\text{pb}}}
\newcommand{\finalResultTotUnctChanSeven}{\ensuremath{67.5\pm5.7\,\,\text{pb}}}
\newcommand{\relPrecisiontChanSeven}{\ensuremath{8.4\%}}
\newcommand{\blueChiSqtChanSeven}{\ensuremath{0.01}}
\newcommand{\blueChiSqProbtChanSeven}{\ensuremath{93}}
\newcommand{\blueOverallCorrProbtChanSeven}{\ensuremath{20}}
\newcommand{\ATLAStChanSevenCoefftChanSeven}{\ensuremath{0.42}}

\newcommand{\CMStChanSevenCoefftChanSeven}{\ensuremath{0.58}}

 \newcommand{\finalResulttChanEight}{\ensuremath{\sigmatch = 87.7\pm1.1\;(\text{stat.})\pm5.5\;(\text{syst.})\pm1.5\;(\text{lumi.})\,\,\text{pb} = 87.7\pm5.8\,\,\text{pb}}}
\newcommand{\finalResultTotUnctChanEight}{\ensuremath{87.7\pm5.8\,\,\text{pb}}}
\newcommand{\relPrecisiontChanEight}{\ensuremath{6.7\%}}
\newcommand{\blueChiSqtChanEight}{\ensuremath{0.59}}
\newcommand{\blueChiSqProbtChanEight}{\ensuremath{44}}
\newcommand{\blueOverallCorrProbtChanEight}{\ensuremath{42}}
\newcommand{\ATLAStChanEightCoefftChanEight}{\ensuremath{0.68}}

\newcommand{\CMStChanEightCoefftChanEight}{\ensuremath{0.32}}

 \newcommand{\finalResulttWSeven}{\ensuremath{\sigmatW = 16.3\pm2.3\;(\text{stat.})\pm3.3\;(\text{syst.})\pm0.7\;(\text{lumi.})\,\,\text{pb} = 16.3\pm4.1\,\,\text{pb}}}
\newcommand{\finalResultTotUnctWSeven}{\ensuremath{16.3\pm4.1\,\,\text{pb}}}
\newcommand{\relPrecisiontWSeven}{\ensuremath{25\%}}
\newcommand{\blueChiSqtWSeven}{\ensuremath{0.01}}
\newcommand{\blueChiSqProbtWSeven}{\ensuremath{91}}
\newcommand{\blueOverallCorrProbtWSeven}{\ensuremath{17}}
\newcommand{\ATLAStWSevenCoefftWSeven}{\ensuremath{0.41}}

\newcommand{\CMStWSevenCoefftWSeven}{\ensuremath{0.59}}

 \newcommand{\finalResulttWEight}{\ensuremath{\sigmatW = 23.1\pm1.1\;(\text{stat.})\pm3.3\;(\text{syst.})\pm0.8\;(\text{lumi.})\,\,\text{pb} = 23.1\pm3.6\,\,\text{pb}}}
\newcommand{\finalResultTotUnctWEight}{\ensuremath{23.1\pm3.6\,\,\text{pb}}}
\newcommand{\relPrecisiontWEight}{\ensuremath{15.6\%}}
\newcommand{\blueChiSqtWEight}{\ensuremath{0.01}}
\newcommand{\blueChiSqProbtWEight}{\ensuremath{94}}
\newcommand{\blueOverallCorrProbtWEight}{\ensuremath{40}}
\newcommand{\ATLAStWEightCoefftWEight}{\ensuremath{0.70}}

\newcommand{\CMStWEightCoefftWEight}{\ensuremath{0.30}}

 \newcommand{\finalResultsChanEight}{\ensuremath{\sigmasch = 4.9\pm0.8\;(\text{stat.})\pm1.2\;(\text{syst.})\pm0.2\;(\text{lumi.})\,\,\text{pb} = 4.9\pm1.4\,\,\text{pb}}}
\newcommand{\finalResultTotUncsChanEight}{\ensuremath{4.9\pm1.4\,\,\text{pb}}}
\newcommand{\relPrecisionsChanEight}{\ensuremath{30\%}}
\newcommand{\blueChiSqsChanEight}{\ensuremath{1.45}}
\newcommand{\blueChiSqProbsChanEight}{\ensuremath{23}}
\newcommand{\blueOverallCorrProbsChanEight}{\ensuremath{15}}
\newcommand{\ATLASsChanEightCoeffsChanEight}{\ensuremath{0.99}}

\newcommand{\CMSsChanEightCoeffsChanEight}{\ensuremath{0.01}}

\newcommand{\totUncvtbComb}{\ensuremath{0.04}}

\newcommand{\finalResultSquaredWithTheoryXSvtbComb}{\ensuremath{|\flv\vtb|^2 = 1.05\pm0.02\;(\text{stat.})\pm0.06\;(\text{syst.})\pm0.01\;(\text{lumi.})\pm0.04\;(\text{theo.}) = 1.05\pm0.08}}

\newcommand{\relPrecisionSquaredvtbComb}{\ensuremath{7.4\%}}

\newcommand{\finalResultWithTheoryXSvtbComb}{\ensuremath{|\flv\vtb|=~&~1.02\pm0.01\;(\text{stat.})\pm0.03\;(\text{syst.})\pm0.01\;(\text{lumi.})\pm0.02\;(\text{theo.}) \\=~&~1.02\pm0.04\;(\text{meas.})\pm0.02\;(\text{theo.}) = 1.02\pm0.04}}
\newcommand{\finalResultWithTheoryXSCompactvtbComb}{\ensuremath{|\flv\vtb| = 1.02\pm0.04\;(\text{meas.})\pm0.02\;(\text{theo.})}}

\newcommand{\relPrecisionvtbComb}{\ensuremath{3.7\%}}

\usepackage{graphicx}
\usepackage{subfig}
\usepackage{colortbl}
\usepackage{multirow}
\usepackage{rotating}
\usepackage{bm}
\usepackage{adjustbox}
\usepackage{dcolumn}
\usepackage{relsize}

\usepackage{caption}
\usepackage{hyphenat}
\allowdisplaybreaks

\newcommand*{\MYPAPERTITLE}{Combinations of single-top-quark production cross-section measurements and \makebold[b]{\absvtb} determinations at {$\boldmath \cmenergyComb$} with the ATLAS and CMS experiments}
\AtlasTitle{\MYPAPERTITLE}

\PreprintIdNumber{CERN-EP-2019-005}

\newcommand*{\MYABSTRACTTEXT}{
This paper presents the combinations of single-top-quark production cross-section measurements by the ATLAS and CMS Collaborations, using data from LHC proton--proton collisions at \cmenergyComb\ corresponding to integrated luminosities of 1.17 to 5.1~\invfb at \cmenergySeven, and  12.2 to 20.3~\invfb at \cmenergyEight. These combinations are performed per centre-of-mass energy and for each production mode: \tch, \myWt, and \sch. 
The combined \tch\ cross-sections are $\finalResultTotUnctChanSeven$ and $\finalResultTotUnctChanEight$ at \cmenergyComb\ respectively. 
The combined \myWt\ cross-sections are $\finalResultTotUnctWSeven$ and $\finalResultTotUnctWEight$ at \cmenergyComb\ respectively.
For the \sch\ cross-section, the combination yields $\finalResultTotUncsChanEight$ at \cmenergyEight.
The square of the magnitude of the CKM matrix element $\vtb$ multiplied by a form factor \flv\ is determined for each production mode and centre-of-mass energy, using the ratio of the measured cross-section to its theoretical prediction. It is assumed that the top-quark-related CKM matrix elements obey the relation $|\vtd|,|\vts| \ll |\vtb|$. All the $\vtbsq$ determinations, extracted from individual ratios at \cmenergyComb, are combined, resulting in \finalResultWithTheoryXSCompactvtbComb. All combined measurements are consistent with their corresponding Standard Model predictions.
}

\newcommand*{\ATLASPAPERNUMBER}{ATLAS TOPQ-2017-16, \\CMS PAS TOP-17-006}

\newcommand*{\MYJOURNAL}{JHEP}

\AtlasRefCode{\ATLASPAPERNUMBER}
\AtlasJournal{\MYJOURNAL}
\AtlasJournalRef{JHEP 05 (2019) 088}
\AtlasDOI{10.1007/JHEP05(2019)088}
\AtlasAbstract{\MYABSTRACTTEXT}
 \AtlasAuthorContributor{The ATLAS and CMS Collaborations}{ }{ }

\hypersetup{pdftitle={ATLAS and CMS document},pdfauthor={The ATLAS and CMS Collaborations}}

\begin{document}

\maketitle

\tableofcontents
\clearpage

\section{Introduction}
\label{intro}

Measurements of single-top-quark production via the electroweak interaction, a process first observed in proton--antiproton (\myppbar) collisions at the Tevatron~\cite{Aaltonen:2009jj, Abazov:2009ii}, have entered the precision era at the Large Hadron Collider (LHC). It has become possible to measure top-quark properties using single-top-quark events~\cite{Giammanco:2017xyn}. Single-top-quark production is sensitive to new physics mechanisms~\cite{Tait:2000sh} that either modify the \mytWb coupling~\cite{Cao:2007ea,Godbole:2011vw,Zhang:2010dr,AguilarSaavedra:2008gt,Dror:2015nkp,Berger:2015yha} or introduce new particles and interactions~\cite{He:1999vp,Aguilar-Saavedra:2013qpa,Durieux:2014xla,Nutter:2012an,Hashemi:2013raa,Drueke:2014pla}. The production rate of single top quarks is proportional to the square of the left-handed coupling at the \mytWb production vertex, assuming that there are no significant \mytWd or \mytWs contributions. In the Standard Model (SM), this coupling is given by the Cabibbo--Kobayashi--Maskawa (CKM)~\cite{CKM1,CKM2} matrix element $\vtb$. Indirect measurements of $|\vtb|$, from precision measurements of $B$-meson decays~\cite{Tanabashi:2018oca} and from top-quark decays~\cite{Abazov:2011zk,Aaltonen:2013luz,Aaltonen:2014yua,CMS-TOP-12-035}, rely on the SM assumptions that the CKM matrix is unitary
and that there are three quark generations. The most stringent indirect determination comes from a global fit to all available $B$-physics measurements, resulting in $|\vtb| = 0.999105 \pm 0.000032$~\cite{Tanabashi:2018oca}. This fit also assumes the absence of any new physics mechanisms that might affect \bquarks. The most precise indirect measurement using top-quark events was performed by the CMS Collaboration in proton--proton (\mypp) collisions at a centre-of-mass energy of \cmenergySeven, resulting in $|\vtb| = 1.007\pm0.016$~\cite{CMS-TOP-12-035}.

A direct estimate of the coupling at the \mytWb production vertex, $\absvtb$, is obtained from the measured single-top-quark cross-section $\sigmameas$ and its corresponding theoretical expectation $\sigmatheo$,
\begin{eqnarray}
\label{eq:vtb}
\absvtb =  \sqrt{\frac{\sigmameas}{\sigmatheo \mathsmaller{\;(\vtb=1)}}}.
\end{eqnarray}
The $\flv$ term is a form factor, assumed to be real, that parameterises the possible presence of anomalous left-handed vector couplings~\cite{AguilarSaavedra:2008zc}. By construction, this form factor is exactly one in the SM, while it can be different from one in models of new physics processes. The direct estimation assumes that $|\vtd|,|\vts| \ll |\vtb|$~\cite{Abazov:2012uga,Alwall:2006bx}, and that the \mytWb interaction involves a left-handed weak coupling, like that in the SM. The $\absvtb$ determination via single-top-quark production is independent of assumptions about the number of quark generations and the unitarity of the CKM matrix~\cite{Tait:1999cf,Belyaev:2000me,Tait:2000sh,Cao:2015qta}. Since the indirect determination of $|\vtb|$ gives a value close to unity, $\vtb$ is considered equal to one in theoretical calculations of the single-top-quark cross-section. The combination of single-top-quark measurements from the Tevatron gives $\absvtb=1.02_{-0.05}^{+0.06}$~\cite{Aaltonen:2015cra}.

\begin{figure}[!h!tbp]
  \begin{center}
    \subfloat[]{ \includegraphics[height=0.12\textheight]{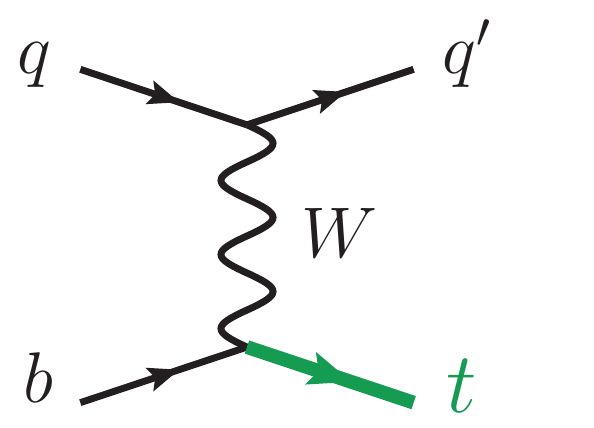} }
    \subfloat[]{ \includegraphics[height=0.12\textheight]{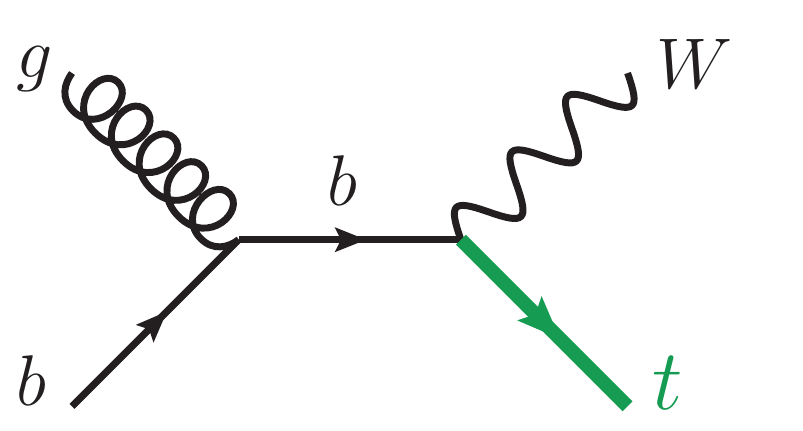} }
    \subfloat[]{ \includegraphics[height=0.12\textheight]{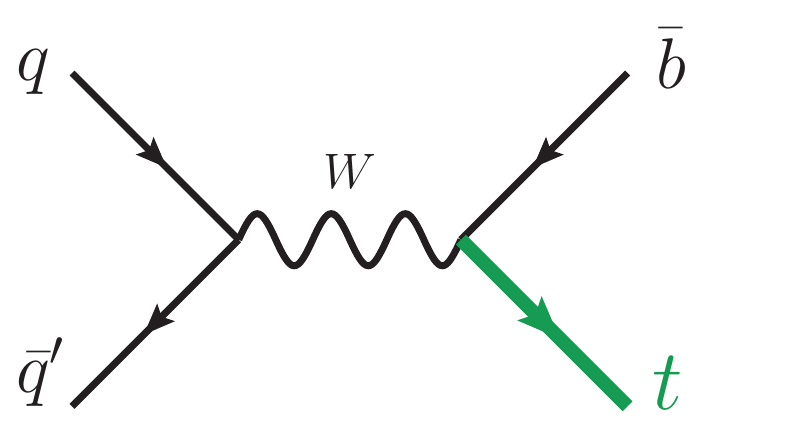} }
    \caption{Representative Feynman diagrams at LO in QCD and in the five-flavour-number scheme for single-top-quark production in (a) the \tch, (b) \myWt production, and (c) the \sch.}
    \label{fig:feyn}
  \end{center}
\end{figure}

Single-top-quark production at a hadron collider mostly proceeds, according to the SM prediction, via three modes that can be defined at leading order (LO) in perturbative quantum chromodynamics (QCD): the exchange of a virtual \myW boson in the \tch\ or in the \sch, and the associated production of a top quark and a \myW boson (\mytW). Representative Feynman diagrams for these processes at LO are shown in Figure~\ref{fig:feyn}.

In \mypp collisions at the LHC, the process with the largest single-top-quark production cross-section is the \tch, where a light-flavour quark \myq from one of the colliding protons interacts with a \bquark by exchanging a space-like virtual \myW boson, producing a top quark (\myt-quark) and a recoiling light-flavour quark $\myq'$, called the spectator quark.
For \tch production at LO, the \bquark can be considered as directly emitted from the other proton (five-flavour-number scheme or 5FS) or it can come from gluon splitting (four-flavour-number scheme or 4FS)~\cite{Frederix:2012dh}. The kinematic properties of the spectator quark provide distinctive features for this process~\cite{TOPQ-2015-05,CMS-TOP-12-038}. The associated production of a \myW boson and a top quark has the second-largest production cross-section. In a representative process of \mytW production, a gluon interacts with an initial \bquark by exchanging a virtual \bquark, producing a \myt-quark and a \myW boson. The measurement of this process suffers from a large background from top-quark pair (\mytt) production~\cite{TOPQ-2012-20,CMS-TOP-12-040}. The \sch\ cross-section is the smallest at the LHC. In this process, a quark--antiquark pair annihilates to produce a time-like virtual \myW boson, which decays to a \myt-quark and a \mybbar-quark. This process was observed in \myppbar collisions at the Tevatron~\cite{CDF:2014uma} 
and evidence of it was reported by the ATLAS Collaboration in \mypp collisions at \cmenergyEight~\cite{TOPQ-2015-01}.

In this paper, the \tch, \myWt, and \sch\ single-top-quark cross-section measurements by the ATLAS and CMS experiments are combined for each production mode, separately at $pp$ centre-of-mass energies of \lhcenergyComb. A combined determination of $\absvtb$ is also presented, using as inputs the values of $\vtbsq$ calculated from the measured and predicted single-top-quark cross-sections in the three production modes at \cmenergyComb. Using the same approach, results are also shown for $\absvtb$ combinations for each production mode.

The theoretical cross-section calculations are described in Section~\ref{sec:theory}. Section~\ref{sec:atlascmsxs} presents the cross-section measurements. The combination methodology is briefly described in Section~\ref{sec:method}. Section~\ref{sec:systcat} is devoted to a discussion of systematic uncertainties in the cross-section measurements as well as theoretical calculations, where the latter affect the $\absvtb$ extraction in particular.
The assumptions made about the correlation of uncertainties between the two experiments, as well as between theoretical calculations, are also discussed. 
Section~\ref{sec:xscomb} presents the combination of cross-sections for each production mode at the same centre-of-mass energy.
In Section~\ref{sec:vtbcomb}, determinations of $\absvtb$ are performed using all single-top-quark cross-section measurements together or by production mode. Stability tests are also shown and discussed. In Section~\ref{sec:sum}, the results are summarised.

\section{Theoretical cross-section calculations}
\label{sec:theory}

The theoretical predictions for the single-top-quark production cross-sections are calculated at next-to-leading order (NLO) in the strong coupling constant $\alpha_{\rm s}$, at NLO with next-to-next-to-leading-logarithm (NNLL) resummation (named NLO+NNLL), and at next-to-next-to-leading order (NNLO). The difference between 4FS and 5FS is small~\cite{Campbell:2009ss, Campbell:2009gj}, and the calculations use the 5FS.
The NLO prediction is used in the \vtb\ combination for the  \tch\ and \sch, while the NLO+NNLL prediction is used for \mytW, as explained below.
The NLO prediction is calculated with \hathor (v2.1)~\cite{Aliev:2010zk,Kant:2014oha}. Uncertainties comprise the scale uncertainty, the $\alpha_{\rm s}$ uncertainty, and the parton distribution function (PDF) uncertainty. The scale uncertainty is evaluated by varying the renormalisation and factorisation scales up and down together by a factor of two. The combination of the PDF+$\alpha_{\rm s}$ uncertainty is calculated according to the PDF4LHC prescription~\cite{Botje:2011sn} from the envelope of the uncertainties at 68\% confidence level (CL) in the MSTW2008 NLO, CT10 NLO~\cite{Lai:2010vv}, and NNPDF2.3~\cite{Ball:2012cx} PDF sets. 

The NLO+NNLL predictions~\cite{Kidonakis:2012rm} are available for all single-top-quark production modes~\cite{Kidonakis:2011wy, Kidonakis:2010ux, Kidonakis:2010tc}. Uncertainties in these calculations are estimated by varying the renormalisation and factorisation scales between $\topmass/2$ and $2 \topmass$, where $\topmass$ is the top-quark mass, and from the 90\% CL uncertainties in the MSTW2008 NNLO~\cite{MSTW, MSTW2008} PDF set. The evaluation of the PDF uncertainties is provided by the author of Refs.~\cite{Kidonakis:2011wy, Kidonakis:2010ux, Kidonakis:2010tc} and is not fully compatible with the PDF4LHC prescription.
The \tch\ cross-sections at \cmenergyComb\ are also computed at NNLO in $\alpha_{\rm s}$~\cite{Brucherseifer:2014ama}, with the renormalisation and factorisation scales set to $\topmass$. This results in cross-sections which are about 0.3\% and 0.6\% lower than the NLO values at \cmenergyComb\ respectively. However, only a limited number of scale variations are evaluated~\cite{Brucherseifer:2014ama}.

A summary of all the available theoretical cross-section predictions for \tch, \mytW, and \sch production, $\sigmatchtheo$, $\sigmatWtheo$, and $\sigmaschtheo$ respectively, with their uncertainties is shown in Table~\ref{tab:xs}.

\begin{table}[!htb]
  \centering
  \caption{Predicted cross-sections for single-top-quark production at \cmenergyComb\ at the LHC. Uncertainties include scale and PDF+$\alpha_{\rm s}$ variations, except for the NNLO predictions, which only contain the scale variation. The PDF+$\alpha_{\rm s}$ uncertainties are evaluated according to the PDF4LHC prescription only for the NLO predictions. The uncertainties associated with the top-quark mass $\topmass$ and beam energy $E_{\rm beam}$ are also given for the NLO predictions for the $t$- and $s$-channels, and for the NLO+NNLL prediction for \mytW\ production. The value of $\topmass$ is set to $172.5~\GeV$ in all predictions. The cross-sections marked with $^\dagger$ are those used in the \absvtb\ combination.}
  \IfPackagesLoaded{adjustbox}{\begin{adjustbox}{max width=1.0\textwidth}}{}
    \begin{tabular}{l|l|l|l}
      \hline
      \hline
      \T\B
      $\sqrt{s}$ & Process & Accuracy & $\sigmatheo{}$\,[pb] \\
      \hline
      \T\B
                 & & NLO${}^{\dagger}$ & $63.9_{-1.3}^{+1.9}~\text{(scale)} \pm 2.2~\text{(PDF+$\alpha_{\rm s}$)} \pm 0.7~\text{($m_\myt$)} \pm 0.1~\text{($E_{\rm beam}$)}$ \\
      \cline{3-4}
      \T\B
                 & \tch & NLO+NNLL & $64.6^{+2.6}_{-1.7}~\text{(scale+PDF+$\alpha_{\rm s}$)}$\\
      \cline{3-4}
      \T\B
                 & & NNLO & $63.7^{+0.5}_{-0.3}~\text{(scale)}$\\
      \cline{2-4}
      \T\B
      \lhcenergySeven & \multirow{2}{*}{\mytW} & NLO & $13.2_{-0.6}^{+0.5}~\text{(scale)} \pm 1.3~\text{(PDF+$\alpha_{\rm s}$)}$ \\
      \cline{3-4}
      \T\B
                 & & NLO+NNLL${}^{\dagger}$ & $15.74 \pm 0.40~\text{(scale)} {}^{+1.10}_{-1.14}~\text{(PDF+$\alpha_{\rm s}$)} \pm 0.28~\text{($m_\myt$)} \pm 0.04~\text{($E_{\rm beam}$)}$\\
      \cline{3-4}
      \cline{2-4}
      \T\B
                 & \multirow{2}{*}{\sch} & NLO${}^{\dagger}$ & $\;\,4.29_{-0.10}^{+0.12}~\text{(scale)} \pm 0.14~\text{(PDF+$\alpha_{\rm s}$)} \pm 0.10~\text{($m_\myt$)} \pm 0.01~\text{($E_{\rm beam}$)}$ \\
      \cline{3-4}
      \T\B
                 & & NLO+NNLL & $\;\,4.63^{+0.20}_{-0.18}~\text{(scale+PDF+$\alpha_{\rm s}$)}$\\
      \cline{3-4}
      \hline
      \hline
      \T\B
                 & & NLO${}^{\dagger}$ & $84.7_{-1.7}^{+2.6}~\text{(scale)} \pm 2.8~\text{(PDF+$\alpha_{\rm s}$)} \pm 0.8~\text{($m_\myt$)} \pm 0.2~\text{($E_{\rm beam}$)}$ \\
      \cline{3-4}
      \T\B
                 & \tch & NLO+NNLL & $87.8^{+3.4}_{-1.9}~\text{(scale+PDF+$\alpha_{\rm s}$)}$\\
      \cline{3-4}
      \T\B
                 & & NNLO & $84.2^{+0.3}_{-0.2}~\text{(scale)}$\\
      \cline{2-4}
      \T\B
                 \lhcenergyEight & \multirow{2}{*}{\mytW} & NLO & $18.77_{-0.82}^{+0.77}~\text{(scale)} \pm 1.70~\text{(PDF+$\alpha_{\rm s}$)}$ \\
      \cline{3-4}
      \T\B
                 & & NLO+NNLL${}^{\dagger}$ & $22.37 \pm 0.60~\text{(scale)} \pm1.40~\text{(PDF+$\alpha_{\rm s}$)} \pm 0.38~\text{($m_\myt$)} \pm 0.06~\text{($E_{\rm beam}$)}$\\
      \cline{3-4}
      \cline{2-4}
      \T\B
                 & \multirow{2}{*}{\sch} & NLO${}^{\dagger}$ & $\;\,5.24_{-0.12}^{+0.15}~\text{(scale)} \pm 0.16~\text{(PDF+$\alpha_{\rm s}$)} \pm 0.12~\text{($m_\myt$)} \pm 0.01~\text{($E_{\rm beam}$)}$ \\
      \cline{3-4}
      \T\B
                 & & NLO+NNLL & $\;\,5.61\pm0.22~\text{(scale+PDF+$\alpha_{\rm s}$)}$\\
      \cline{3-4}
      \hline
      \hline
    \end{tabular}
    \IfPackagesLoaded{adjustbox}{\end{adjustbox}}{}
  \label{tab:xs}
\end{table}

In this paper, NLO predictions serve as the reference for the $t$- and $s$-channel processes, following the prescriptions presented above, because higher-order calculations and their uncertainties are not fully computed and available for the parameter values of choice. The advantage of the NLO cross-section calculations is that the configurable parameters in \hathor can be set according to those used to generate the ATLAS and CMS simulation samples. The $t$- and \sch\ processes do not interfere at NLO~\cite{Willenbrock:1986cr}. For these two processes, the entire phase space is included in the integration in order to obtain the total cross-section. The \mytW cross-section prediction, $\sigmatWtheo$, is available at NLO ~\cite{Kant:2014oha} and NLO+NNLL~\cite{Kidonakis:2013zqa,Kidonakis:2010ux}. The \mytW process at NLO interferes with the \mytt process at LO with the subsequent decay $\mytbar \rightarrow W\mybbar$. In the NLO prediction for \mytW production provided in Ref.~\cite{Kant:2014oha}, a kinematic cut-off is imposed on the transverse momentum (\pt) of the outgoing \bquark, suppressing the contribution from \ttbar\ production. Since the treatment of this interference in \hathor is still being developed~\cite{Demartin:2016axk, Campbell:2005bb}, the NLO+NNLL calculation is used as reference for \mytW production. For the reference cross-section predictions, uncertainties corresponding to the dependence on $\topmass$ and on the LHC beam energy, $E_{\rm beam}$, are evaluated. The $\topmass$ dependence is estimated by varying its central value of $172.5~\GeV$ (the value used in the simulation samples used to measure the single-top-quark cross-sections) by $\pm$1~\GeV, using the functional form proposed in Ref.~\cite{Czakon:2013goa}. The theoretical calculations are performed at a given centre-of-mass energy while the energy of the LHC beam is measured with an uncertainty. The single-top-quark cross-sections are assumed to depend on $E_{\rm beam}$ according to the model given in Ref.~\cite{Langenfeld:2009tc}, with a relative uncertainty $\delta E_{\rm beam}/E_{\rm beam}$ of 0.1\%~\cite{Todesco:2017nnk}. 
The theoretical cross-sections that are used as reference are marked with a $^\dagger$ in Table~\ref{tab:xs}.

 \section{Single-top-quark cross-section measurements at \cmenergyComb}
\label{sec:atlascmsxs}

The \tch single-top-quark production cross-sections, $\sigmatch$, were measured by the
ATLAS and CMS Collaborations at \cmenergySeven~\cite{TOPQ-2012-21,CMS-TOP-11-021}
and 8\,\TeV~\cite{TOPQ-2015-05,CMS-TOP-12-038}. Evidence of \mytW production was reported at \cmenergySeven~by 
ATLAS~\cite{TOPQ-2011-17} and CMS~\cite{CMS-TOP-11-022}, 
while at \cmenergyEight\ its cross-section, $\sigmatW$, was
measured by both
experiments~\cite{TOPQ-2012-20,CMS-TOP-12-040}. Evidence of \sch\
production was reported by ATLAS, with a measured cross-section,
$\sigmasch$, at \cmenergyEight~\cite{TOPQ-2015-01}, whereas CMS set upper limits on
the \sch\ production cross-section at \cmenergyComb. The observed
(expected) significance of the CMS measurement at \cmenergyEight\ is $2.3$ ($0.8$) standard
deviations~\cite{CMS-TOP-13-009}.

The ATLAS and CMS analyses use similar approaches to measure the single-top-quark production cross-sections. Both experiments select events containing at least one prompt isolated lepton (electron or muon) and at least one high-\pt\ jet. The analyses use various multivariate analysis (MVA) techniques, such as boosted decision trees~\cite{FREUND1997119,Friedman2006,Hocker:2007ht}, neural networks~\cite{Feindt:2006pm}, or the matrix element method (MEM)~\cite{Kondo:1988yd, Kondo:1991dw}, to separate the signal from background. To measure the cross-section, analyses perform a binned maximum-likelihood fit to data using the distribution of the corresponding MVA discriminator. 
Exceptions are the ATLAS \sch\ and CMS \tch\ measurements at \cmenergyEight. In the ATLAS \sch\ analysis, the fit is performed simultaneously to the MEM discriminant in the signal region and the lepton-charge distribution in the \myWjets control region. The CMS \tch\ measurement at \cmenergyEight\ is based on a simultaneous fit to the absolute pseudorapidity ($\eta$) distributions of the recoiling light-flavour jet in events with negative and with positive lepton charge. 
The analyses measuring different single-top-quark production modes within the same experiment and at the same centre-of-mass energy have disjoint signal regions. 
Both experiments simulate the single-top-quark processes using the NLO \POWHEGBOX\ generator~\cite{Nason:2004rx,Frixione:2007vw,Alioli:2009je,Alioli:2010xd,Re:2010bp} for the matrix-element (ME) calculations. ATLAS also uses the \POWHEGBOX\ generator to simulate top-quark-pair background events, while CMS uses the LO \MADGRAPH generator~\cite{Alwall:2011uj}. The \PYTHIA~\cite{Sjostrand:2006za} event generator is used for modelling the parton shower (PS), hadronisation and the underlying event in both the single-top-quark and \ttbar\ processes. The cross-sections are measured assuming a value of $172.5$~\GeV\ for $\topmass$ for all top-quark processes
and all centre-of-mass energies. 
A summary of the uncertainties in each measurement is shown in Table~\ref{tab:xsec_meas}, with details given in Appendix~\ref{app:allUncs}.

\begin{table}[!htbp]
  \centering
  \caption{Summary of the single-top-quark cross-section measurements published by the ATLAS and
    CMS Collaborations at
    \cmenergyComb. Total uncertainties are shown. Small
    differences between the integrated luminosity 
    values in different analyses within the same experiment and
    centre-of-mass energy are due to different luminosity
    calibrations at the time of publication.}
  \IfPackagesLoaded{adjustbox}{\begin{adjustbox}{max width=1.0\textwidth}}{}
    \begin{tabular}{l|l|r@{\hskip 0cm}lc|r@{\hskip 0cm}lc}
      \hline
      \hline
      \multicolumn{2}{c|}{} & \multicolumn{3}{c|}{ATLAS} & \multicolumn{3}{c}{CMS} \\
      \hline
      $\sqrt{s}$ & Process & \multicolumn{2}{c}{$\sigma{}$\,[pb]} & Lumi.\ [$\invfb$] & \multicolumn{2}{c}{$\sigma{}$\,[pb]} & Lumi.\ [$\invfb$] \\
      \hline
                            & \tch & $68$&$~\pm~8$ & 4.59 & $67.2$&$~\pm~6.1$ & 1.17--1.56 \\ 
      \cline{2-8}
      \T\B
      \lhcenergySeven & \mytW & $16.8$&$~\pm~5.7$ & 2.05 & \multicolumn{2}{c}{$16^{+5}_{-4}$} & 4.9 \\
      \cline{2-8}
      \T\B
                            & \sch & \multicolumn{2}{c}{---} & --- & $7.1$&$~\pm~8.1$ & 5.1 \\
      \hline
      \hline
      \T\B
                            & \tch & \multicolumn{2}{c}{$89.6^{+7.1}_{-6.3}$} & 20.2 & $83.6$&$~\pm~7.8$ & 19.7 \\
      \cline{2-8}
      \T\B
      \lhcenergyEight & \mytW & \multicolumn{2}{c}{$23.0^{+3.6}_{-3.9}$} & 20.3 & $23.4$&$~\pm~5.4$ & 12.2 \\
      \cline{2-8}
      \T\B
                            & \sch & \multicolumn{2}{c}{$4.8^{+1.8}_{-1.5}$} & 20.3 & $13.4$&$~\pm~7.3$ & 19.7 \\
      \hline
      \hline
    \end{tabular}
    \renewcommand{\arraystretch}{1}
    \IfPackagesLoaded{adjustbox}{\end{adjustbox}}{}
  \label{tab:xsec_meas}
\end{table}
  \section{Combination methodology}
\label{sec:method}
The ATLAS and CMS single-top-quark production cross-section measurements shown in Table~\ref{tab:xsec_meas} are combined, and the combined $\absvtb$ value determined, using the best linear unbiased estimator (BLUE) method~\cite{Lyons:1988rp, Valassi:2003mu,Nisius:2014wua}. The BLUE method is applied iteratively in order to reduce a possible bias arising from the dependence of systematic uncertainties on the central value of the cross-section~\cite{Lista:2014qia}. Convergence is reached when the central value changes by less than 0.01\% compared with the previous iteration. 
In each iteration, the BLUE method minimises the global $\chi^2$ by adjusting the weight for each input measurement~\cite{,Nisius:2014wua}. The global $\chi^2$ is calculated taking correlations
into account. The sum of weights is required to be equal to one. Negative weights are allowed; these indicate strong correlations~\cite{Valassi:2013bga}. The number of degrees of freedom is $n-1$, where $n$ is number of measurements in the combination. The $\chi^2$ and $n$ are then used to calculate a corresponding probability~\cite{Nisius:2014wua}. The systematic uncertainties are scaled with the cross-section in each iteration, i.e.\ they are treated as relative uncertainties. The data and simulation statistical uncertainties are not scaled~\cite{Lista:2014qia}. The systematic uncertainties in the \sch\ cross-section combination are also not scaled because the \sch\ measurements have large backgrounds.

Following the same strategy as in the input measurements by the ATLAS and CMS Collaborations, the combined cross-sections are reported at $\topmass = 172.5~\GeV$, not including the uncertainty associated with the $\topmass$ variation. The shift in the combined cross-section due to a variation of $\pm 1~\GeV$ in the top-quark mass is given where this information is available. For the determination of the combined $\absvtb$ value, the uncertainty in the measured cross-sections due to a variation of $\pm 1~\GeV$ in the mass is considered.
Uncertainties in the measurements are symmetrised, before combination, by averaging the magnitude of the downward and upward variations. More details are given in Sections~\ref{sec:systcat} and \ref{sec:xscomb}.
 \section{Systematic uncertainties and correlation assumptions}
\label{sec:systcat}
In order to combine single-top-quark cross-section measurements and $\absvtb$ values, the sources of uncertainty are grouped into categories. While the categorisation and evaluation of uncertainties varies somewhat between experiments and between measurements, each individual measurement considers a complete set of uncertainties. Assumptions are made about correlations between similar sources of uncertainty in different measurements, as explained in Section~\ref{sec:systcat_exp}. Uncertainties associated with theoretical predictions are taken into account in the $\absvtb$ combination. The correlations between similar uncertainties in different theoretical predictions are discussed in Section~\ref{sec:systcat_theo}.

\subsection{Systematic uncertainties in measured cross-sections}
\label{sec:systcat_exp}

Systematic uncertainties in the ATLAS \tch\ measurements at \cmenergyComb\ are evaluated using pseudoexperiments, except the background normalisation uncertainties, which are constrained in the fit to data. In the ATLAS \mytW\ measurements at \cmenergyComb\ and the \sch\ measurement at \cmenergyEight, systematic uncertainties are included as nuisance parameters in profile-likelihood fits. Systematic uncertainties in the CMS \tch\ and \mytW\ measurements at \cmenergyComb\ are included as nuisance parameters in fits to data, except the theory modelling uncertainties in signal and backgrounds, described below, which are evaluated using pseudoexperiments. All systematic uncertainties in the CMS \sch\ measurements at \cmenergyComb\ are obtained through pseudoexperiments, except the background normalisation uncertainties, which are constrained in the fit to data. In the analyses where systematic uncertainties are included as nuisance parameters, the total uncertainty presented in Table~\ref{tab:xsec_meas} is evaluated by varying all the nuisance parameters in the fit simultaneously. To extract the impact of each source of this type of uncertainty, these analyses use approximate procedures which neglect the correlations between sources of uncertainty introduced by the fits. Throughout this paper, individual uncertainties are taken as reported by the input analyses, regardless of the method used to determine them. The total uncertainties are evaluated as the sum in quadrature of individual contributions.

Although the sources of systematic uncertainty and the procedures used to
estimate their impact on the measured cross-section are partially different in the individual analyses, it is still possible to identify contributions that describe similar physical effects. These contributions are listed below; they are grouped together, and only the resulting categories are used in the combination. Categories are treated as uncorrelated among each other. 
For each source of uncertainty, correlations between different measurements are assumed to be positive, unless explicitly mentioned otherwise. 
The stability of the cross-section and $\absvtb$ combinations is studied by varying the correlation assumptions for the dominant uncertainties, as discussed in Section~\ref{sec:tests}.

The uncertainties in each category are listed below, with the correlation assumptions across experiments given in parentheses. These correlations correspond to those used in the cross-section combinations. They are also valid for the combination of the $\absvtb$ extractions, unless explicitly mentioned otherwise. The symbol ``---'' means that the uncertainty is either considered only in the ATLAS or the CMS measurement, or is not considered at all. A summary of uncertainties in the cross-section measurements together with the corresponding correlation assumptions between experiments is provided in Appendix~\ref{app:allUncs}.

\noindent {\bf Data statistical} (Correlation 0)\\ This statistical uncertainty arises from the limited size of the data sample. It is uncorrelated between ATLAS and CMS, between production modes, and between centre-of-mass energies.

\noindent {\bf Simulation statistical} (Correlation 0 and --- for CMS \mytW\ at \cmenergySeven\ and \sch\ at \cmenergyEight)\\ This statistical uncertainty comes from the limited size of simulated event samples. It is uncorrelated between ATLAS and CMS, between production modes, and between centre-of-mass energies. For the CMS \mytW\ analysis at \cmenergySeven\ and \sch\ analysis at \cmenergyEight, this uncertainty is evaluated as part of the total statistical uncertainty, which is also considered uncorrelated, as discussed above. 
More details are given in Appendices~\ref{app:tw} and \ref{app:sch}.

\noindent {\bf Integrated luminosity} (Correlation 0.3)\\ This uncertainty originates from the systematic uncertainty in the integrated luminosity, as determined by the individual experiments using the methods described in Refs.~\cite{DAPR-2011-01,DAPR-2013-01,CMS-PAS-SMP-12-008,CMS-PAS-LUM-13-001}. It affects the determination of both the signal and background yields. The integrated-luminosity uncertainty has a component that is correlated between ATLAS and CMS, arising from imperfect knowledge of the beam currents during van der Meer scans in the LHC accelerator~\cite{vanderMeer:1968zz}, and an uncorrelated component from the long-term luminosity monitoring that is experiment-specific.
At \cmenergySeven, these components are 0.5\% and 1.7\% respectively for ATLAS and 0.5\% and 2.1\% respectively for CMS. At \cmenergyEight, they are 0.6\% and 1.8\% respectively for ATLAS and 0.7\% and 2.5\% respectively for CMS.
At both centre-of-mass energies, the correlation coefficient between the integrated-luminosity uncertainty in ATLAS and CMS at the same centre-of-mass energy is $\rho=0.3$. Within the same experiment, the integrated-luminosity uncertainty is assumed to be correlated between production modes and uncorrelated between centre-of-mass energies. In Section~\ref{sec:tests}, it is shown that the combined $\vtbsq$ result does not depend significantly on the correlation assumptions.

\noindent {\bf Theory modelling}\\ This category contains the uncertainties in the modelling of the simulated single-top-quark processes, as well as smaller contributions from the modelling of the \mytt and \myWjets background processes. Both signal and background modelling are included because the uncertainties in all top-quark processes are closely related.
 These include initial- and final-state radiation (ISR/FSR), renormalisation and factorisation scales, NLO matching method, PS and hadronisation modelling, and PDF uncertainties. For the \mytW process, the uncertainty due to the treatment of interference between \myWt and \mytt final states is also included, as discussed below. These modelling uncertainties in signal and background processes are summed in quadrature in each input measurement.

\begin{itemize}

\item {\it Scales and radiation modelling} (Correlation 1)\\ The renormalisation and factorisation scales and ISR/FSR uncertainties account for missing higher-order corrections in the perturbative expansion and the amount of initial- and final-state radiation in simulated signal and background processes. In the ATLAS measurements of all three production modes, these uncertainties are estimated using dedicated single-top-quark and \mytt simulated event samples, by consistently varying the renormalisation and factorisation scales and the amount of ISR/FSR in accordance with a measurement of additional jet activity in \mytt events at \cmenergySeven~\cite{TOPQ-2011-21,ATL-PHYS-PUB-2014-005}.
In the ATLAS \tch\ measurements, they are also estimated in \myWjets\ simulated event samples, by varying the scale and matching parameters in the \ALPGEN\ LO multileg generator~\cite{Mangano:2002ea} at \cmenergySeven\ and by varying the parameters controlling the scale in the \SHERPA\ LO multileg generator~\cite{Gleisberg:2008ta} at \cmenergyEight. In the CMS measurements, these uncertainties are estimated by varying the renormalisation and factorisation scales, and ISR/FSR, consistently in the simulated event samples. In the CMS \tch\ measurement at \cmenergyEight, this uncertainty applies only to the signal modelling since the modelling of the dominant \mytt and \myWjets background processes is obtained from data. However, for the \tch\ analysis at \cmenergySeven, the scales are varied in the simulated signal, \mytt, \myWjets and other single-top-quark processes. The same approach is followed in the CMS \sch\ measurements at both centre-of-mass energies. The \mytW cross-section measurements of CMS account for this uncertainty only in the \mytW signal and \mytt background, given the negligible contributions from the \myWjets and other single-top-quark processes in the dilepton final state.

Although the methods are apparently different, they mostly address the same uncertainty, hence this uncertainty is considered correlated between ATLAS and CMS. It is also considered correlated between production modes and centre-of-mass energies. The combined $\absvtb$ result does not depend significantly on this correlation assumption,
as discussed in Section~\ref{sec:tests}.

\item {\it NLO matching} (Correlation 1 for \tch\ and --- for \mytW and \sch)\\ The ATLAS measurements include an uncertainty to account for different NLO matching methods implemented in different NLO event generators. This is evaluated in single-top-quark and \mytt simulations by comparing the \POWHEGBOX, \MCatNLO~\cite{Frixione:2002ik,Skands:2010ak}, and \MGMCatNLO~\cite{Alwall:2014hca} event generators, all interfaced to \HERWIG~\cite{Corcella:2000bw} (with \JIMMY~\cite{Butterworth:1996zw} for the underlying-event modelling). In the CMS \tch\ measurement at \cmenergySeven, the NLO matching uncertainty is evaluated by comparing \POWHEGBOX\ with \Comphep~\cite{Boos:2004kh,Pukhov:1999gg}. In the CMS \tch\ analysis at \cmenergyEight, this uncertainty accounts for different NLO matching methods in the \tch\ signal event generator, as well as for differences between event generation in the 4FS and 5FS, by comparing \POWHEGBOX\ with \MADGRAPH. The NLO matching uncertainty is considered correlated between ATLAS and CMS, between production modes, and between centre-of-mass energies. In the CMS \mytW and \sch\ analyses at \cmenergyComb, this uncertainty is not considered, since the modelling uncertainties in the scheme to remove overlap with \mytt are dominant in the \mytW analysis and the renormalisation/factorisation scale is dominant in the \sch\ analysis. The results of the stability test for this uncertainty are shown in Section~\ref{sec:tests}.

\item {\it Parton shower and hadronisation} (Correlation 1)\\ In both experiments, the difference between the \PYTHIA and \HERWIG showering programs is considered in the jet energy scale (JES)~\cite{PERF-2011-03,PERF-2012-01,CMS-JME-10-011,CMS-JME-13-004} and \btag calibration~\cite{PERF-2012-04,ATLAS-CONF-2014-004,ATLAS-CONF-2014-046,CMS-BTV-12-001,CMS-PAS-BTV-13-001}. 
The ATLAS analyses additionally  include an uncertainty in the PS and hadronisation modelling in simulated single-top-quark and \mytt events, evaluated by comparing the \POWHEGBOX\ event generator interfaced to \PYTHIA or to \HERWIG. 
The CMS analyses additionally include an uncertainty in the \mytt and \myWjets backgrounds estimated with the \MADGRAPH event generator interfaced to \PYTHIA. It is evaluated in simulated event samples where the value of the ME/PS matching threshold in the MLM method~\cite{Mangano:2006rw} is doubled or halved from its initial value. 
The CMS \tch\ measurement at \cmenergyEight\ does not consider this uncertainty in the \mytt and \myWjets backgrounds since the distribution and normalisation of the \mytt\ and \myWjets\ processes are derived mostly from data.
In the CMS \mytW analyses at \cmenergyComb, the contributions of the \myWjets and other single-top-quark processes in the dilepton final state are negligible.

This uncertainty is considered correlated between ATLAS and CMS, between different production modes, and between different centre-of-mass energies. The combined $\absvtb$ result does not depend significantly on this correlation assumption, as shown in Section~\ref{sec:tests}.

\item {\it Parton distribution functions} (Correlation 1)\\ The PDF uncertainty is evaluated following the PDF4LHC procedures~\cite{Alekhin:2011sk,Botje:2011sn,Butterworth:2015oua} and is 
considered correlated between ATLAS and CMS, between different production modes, and between different centre-of-mass energies.

\item {\it \mytW and \mytt interference} (Correlation 1 for \mytW\ and --- for $t$- and $s$-channels)\\ The \mytW process interferes with \mytt production at NLO~\cite{Frixione:2008yi, tWth2, tWth3}. In both ATLAS and CMS, two simulation approaches are compared: diagram removal (DR)~\cite{Frixione:2008yi} and diagram subtraction (DS)~\cite{Frixione:2008yi,Tait:1999cf}.
In the DR approach, all NLO diagrams that overlap with the doubly resonant \mytt\ contributions are removed from the calculation of the \mytW\ amplitude. This approach accounts for the interference term, but it is not gauge invariant (though the effect is numerically negligible)~\cite{Frixione:2008yi}. In the DS approach, a subtraction term is built into the amplitude to cancel out the \mytt\ component close to the top-quark resonance while respecting gauge invariance.

 The DR approach is the default, and the comparison with the DS approach is used to assess this systematic uncertainty. For the \mytW analyses, this uncertainty is considered correlated between the two experiments and between different centre-of-mass energies.

\item {\it Modelling of  the top-quark \mypT\ spectrum} (Correlation ---)\\ In the CMS \mytW and \sch\ analyses at \cmenergyEight, the simulated \mytt events are reweighted to correct the $\mypT$\ spectrum of the generated top quarks, which was found to be significantly harder than the spectrum observed in data in differential cross-section measurements~\cite{CMS-TOP-12-028,Czakon:2016ckf}. To estimate the uncertainty related to this mismodelling, the \mytW\ measurement is repeated without the reweighting, and the change relative to the default result is taken as the uncertainty. In the CMS \sch\ analysis, the measurement is repeated with the effect of the weights removed and doubled. The resulting variation in the cross-section is symmetrised. This uncertainty is not considered in the CMS \tch\ measurement at \cmenergyEight\ where the modelling of the \mytt\ background is extracted from data. In the ATLAS measurements, modelling uncertainties in the top-quark \mypT spectrum in \mytt events~\cite{TOPQ-2015-06} are covered by the PS and hadronisation uncertainty and they are found to be small in comparison with other systematic uncertainties. This uncertainty is considered correlated between the CMS \mytW and \sch\ analyses at \cmenergyEight.

\item {\it Dependence on the top-quark mass} (Correlation 1)\\ The measured single-top-quark cross-sections shown in Table~\ref{tab:xsec_meas} assume a nominal $\topmass$ value of 172.5~\GeV. The dependence of the measured cross-section on $\topmass$ is estimated for the ATLAS \tch\ measurements at \cmenergyComb\ and for the ATLAS \mytW\ measurement at \cmenergyEight. It is determined using dedicated simulations of single-top-quark and \ttbar\ samples with different $m_t$ values. The cross-section measurements assuming the different $m_t$ values are interpolated using a first- or a second-order polynomial, for which the constant term is given by the central value of $\topmass = 172.5$~\GeV. 
  The CMS measurements at \cmenergyEight\ provide information for a variation of $\pm 2~\GeV$ in the top-quark mass, which is scaled to a $\pm 1~\GeV$ shift assuming a linear dependence. For the CMS \tch\ and \mytW\ measurements at \cmenergyEight, the changes in cross-sections are symmetrised and reported as uncertainties.  In the CMS \sch\ analysis, the change in the cross-section is determined for the up and down variation of $\topmass$.
  No estimates are available for the CMS \tch\ analysis at \cmenergySeven, the ATLAS and CMS \mytW\ analyses at \cmenergySeven\ or the ATLAS \sch\ analysis at \cmenergyEight. The top-quark-mass uncertainty is small for each measurement, thus the impact of not evaluating it for these measurements is negligible.

 In this paper, a symmetrised uncertainty in the measured cross-section due to a variation of $\pm 1~\GeV$ in the top-quark mass is considered. When the full cross-section dependence on the top-quark mass is available for a given production mode at a given centre-of-mass energy, the sign of the dependence of the uncertainty per unit of mass is taken into account in the correlations.
In the case of the CMS \tch\ and \mytW\ measurements at \cmenergyEight, where the sign of the dependence is not available, it is assumed that the sign is the same as for the ATLAS measurement, since the phase space and background composition are comparable between CMS and ATLAS. Given that the uncertainty in the measured cross-section is considered for the same $m_t$ variation and considering the sign of the dependence when available, this uncertainty is considered correlated between ATLAS and CMS and between different centre-of-mass energies and uncorrelated between the \tch\ and \mytW\ production modes. 

\end{itemize}

\noindent {\bf Background normalisation} (Correlation 0)\\
Three background uncertainties are considered: in top-quark background (\ttbar\ and other single-top-quark processes), in other background determined from simulation (\myWZjets, diboson, and other smaller background channels), and in background estimated from data (multijet background from misidentified and non-prompt leptons). The exceptions are the \tch\ measurements at \cmenergySeven, where the background from simulation includes top-quark background, as shown in Tables~\ref{tab:tChan7XS}$-$\ref{tab:sChan8XS} in Appendix~\ref{app:allUncs}. 
The normalisation of the main background processes is determined from data, either by inclusion of normalisation uncertainties as nuisance parameters in the fit used to extract the signal, or through dedicated techniques based on data. In the \tch\ and \sch\ measurements, the uncertainties in the theoretical cross-section predictions for the top-quark, \myWZjets, and diboson processes are included. In the \mytW\ measurements, the uncertainties in the theoretical cross-section predictions for the top-quark and diboson processes are taken into account. In the ATLAS measurements of the \tch\ process at \cmenergyComb, the uncertainty in the multijet background is estimated by comparing background estimates made using different techniques based on simulation and data samples.
In the ATLAS \mytW\ analyses at \cmenergyComb, the normalisation uncertainty in the background from misidentified and non-prompt leptons is obtained from variations in the data-based estimate.
In the ATLAS \sch\ analysis, the uncertainty assigned to the normalisation of the multijet background is based on control samples.
For all CMS measurements, background normalisations are constrained in the fits to data. In the CMS measurements of the \tch\ and \sch\ processes, the uncertainties in the multijet background are assessed by comparing the results of alternative background estimation methods based on data. 
Hence, the associated uncertainties are considered uncorrelated between ATLAS and CMS, between different production modes, and between different centre-of-mass energies.

\noindent {\bf Jets} \\ In the analyses, the uncertainties related to the reconstruction and energy calibration of jets are propagated through variations in the modelling of the detector response. These uncertainties, classified in categories as JES, jet identification (JetID), and jet energy resolution (JER), are discussed below.

\begin{itemize}
\item {\it Jet energy scale} (Correlation 0 and --- for JES flavour)\\ 
The JES is derived using information from data and simulation. Its uncertainty increases with increasing $|\eta|$ and decreases with increasing $\pT$ of the reconstructed jet.

For all of the ATLAS measurements, except the $\mytW$ measurement at \cmenergySeven, the JES uncertainty is split into  components originating from the jet calibration procedure; most of them are derived from in situ techniques based on data~\cite{PERF-2011-03,PERF-2012-01}. These components are categorised as modelling, detector, calibration method, and statistical components, which are grouped into the ``JES common'' uncertainty, as well as a flavour-dependence component (``JES flavour''),  which accounts for the flavour composition of the jets and the calorimeter response to jets of different flavours. The modelling of additional $pp$ collisions in each bunch-crossing (pile-up) is considered separately, as discussed below. The  $\eta$-dependent component is dominant for the \tch\ production mode. Thus, the JES common uncertainty is considered uncorrelated between the \tch\ and the other single-top-quark production modes. For the \mytW analysis at \cmenergyEight, the modelling component, which is constrained in the fit to data, is dominant. The uncertainty in the flavour composition of the jets is dominant for the \sch. 

For the CMS measurements, sources contributing to the JES uncertainty are combined together into the ``JES common'' uncertainty, and the effect is propagated to the cross-section measurements through $\eta$- and $\mypT$-dependent JES uncertainties~\cite{CMS-JME-10-011,CMS-JME-13-004}. The jet energy corrections and their corresponding uncertainties are extracted from data. The  JES uncertainty is estimated from its effect on the normalisation and shape of the discriminant in each analysis. The JES uncertainty is considered uncorrelated between the \tch\ and the other single-top-quark production modes because it is dominated by the forward jet in the \tch.

The correlation between the JES common uncertainty (or the JES uncertainty for the $\mytW$ measurement at \cmenergySeven) in ATLAS and the JES uncertainty in CMS follows the prescription in Refs.~\cite{ATL-PHYS-PUB-2014-020, ATL-PHYS-PUB-2015-049}, with the slight differences for the \tch\ described above. The JES common (or JES) uncertainty is considered uncorrelated between ATLAS and CMS, between centre-of-mass energies, and between production modes. Within the ATLAS experiment, the JES common uncertainty is considered correlated between \mytW\ and \sch\ and uncorrelated between \tch\ and the other production modes. For the ATLAS \tch\ analyses, a correlation of 0.75 is assumed between \cmenergyComb, since these analyses are mainly affected by the same uncertainty components. This correlation value is estimated by comparing variations of the JES uncertainty components in these two measurements.

In all CMS measurements and in the ATLAS $\mytW$ measurement at \cmenergySeven, the JES uncertainty is not split and therefore the JES flavour uncertainty is included in the overall JES uncertainty. For the ATLAS measurements where this component is available, the JES flavour uncertainty is considered correlated between different production modes and uncorrelated between centre-of-mass energies.

The JES uncertainty is one of the dominant contributions in most of the single-top-quark measurements. To ensure the robustness of the results against the correlation assumptions for this large uncertainty, the combination is performed with alternative correlation values, as discussed in Section~\ref{sec:tests}.

\item {\it Jet identification} (Correlation ---)\\ In the ATLAS measurements, the JetID uncertainty includes the jet and vertex reconstruction efficiency uncertainties. In the CMS measurements, this uncertainty is included in the JES uncertainty.
  For ATLAS, it is considered correlated between the different production modes at the same centre-of-mass energy and uncorrelated for the other cases.
  
\item  {\it Jet energy resolution} (Correlation 0)\\ The uncertainty in the JER, which is not split into components, is extracted from data. Generally, the JER uncertainty is propagated via a nuisance parameter in the signal extraction fit, except for the ATLAS \tch\ measurements at \cmenergyComb, and the CMS \sch\ measurement, where this uncertainty is determined using pseudoexperiments. The JER uncertainty is considered uncorrelated between ATLAS and CMS, and between centre-of-mass energies. It is considered correlated between different production modes.
\end{itemize}

\noindent {\bf Detector modelling}  \\ This category includes the uncertainty in the modelling of leptons, magnitude of the missing transverse momentum (\myMET), and identification of jets from $b$-quarks (\btag).

\begin{itemize}
\item {\it Lepton modelling} (Correlation 0)\\ 
The lepton modelling uncertainty includes components associated with the lepton energy scale and resolution, reconstruction and trigger efficiencies. This uncertainty is considered uncorrelated between ATLAS~\cite{PERF-2014-05,PERF-2013-05,PERF-2013-03,PERF-2016-01} and CMS~\cite{Sirunyan:2017ulk} and between different centre-of-mass energies, since it is determined from data. It is considered correlated between different production modes.

\item {\it Hadronic part of the high-level trigger} (Correlation ---)\\ 
In the CMS \tch\ cross-section measurement at \cmenergySeven, the high-level trigger (HLT) criteria for the electron channel are based on the presence of an electron together with a \btagged jet. In this analysis, the uncertainty in the modelling of the hadronic part of the HLT requirement is determined from data. This uncertainty is only evaluated in this one measurement.

\item {\it \myMET modelling} (Correlation 0)\\ 
The ATLAS measurements include separate components for the uncertainties in the energy scale and resolution of the \myMET~\cite{PERF-2014-04}. The CMS measurements account for a combined \myMET scale and resolution uncertainty~\cite{CMS-JME-10-011,CMS-JME-13-003}, arising from the jet-energy uncertainties.
Additionally, CMS accounts for an uncertainty in \myMET arising from energy deposits in the detector that are not included in the reconstruction of leptons, photons, and jets.
The \myMET uncertainty is considered uncorrelated between ATLAS and CMS, and between different centre-of-mass energies. It is considered correlated between production modes, except for the ATLAS and CMS \mytW analyses at \cmenergyEight, where it is considered  uncorrelated with the other production modes because the \myMET uncertainty is constrained in the fit to data. In the ATLAS \mytW analysis at \cmenergySeven, this uncertainty is included in the pile-up modelling uncertainty.

\item {\it \btag} (Correlation 0)\\ 
In the ATLAS analyses, \btag modelling uncertainties are split into components associated with \bquark, \myc-quark, and light-flavour quark and gluon jets~\cite{PERF-2012-04,ATLAS-CONF-2014-004,ATLAS-CONF-2014-046}. They are evaluated by varying the $\pT$-dependence ($\eta$-dependence in the case of light-flavour jets) of the flavour-dependent scale factors applied to each jet in simulation within a range that reflects the systematic uncertainty in the measured tagging efficiency and misidentification rates. This uncertainty is not considered in the ATLAS \mytW\ analysis at \cmenergySeven\ because no \btag criterion is applied in the event selection. 
In the CMS measurements, the uncertainties in \btag efficiency and misidentification rates of jets initiated by light-flavour quarks and gluons are derived from data, using control samples~\cite{CMS-BTV-12-001,CMS-PAS-BTV-13-001}. The CMS uncertainties are propagated to the cross-section measurements using pseudoexperiments. Exceptions are the \tch\ measurement at \cmenergySeven\ and the \mytW measurement at \cmenergyEight, where these uncertainties are constrained in the fit to data.

The two collaborations split up the different sources of systematic uncertainties related to \btag in a different way. However, the different sources are combined by adding their contributions in quadrature to obtain a single \btag uncertainty per analysis. This means that the \btag uncertainty also contains the uncertainties associated with the misidentification rates of jets initiated by charm quarks, light-flavour quarks and gluons. The resulting uncertainty is considered uncorrelated between ATLAS and CMS, and between different centre-of-mass energies. It is considered correlated between different production modes.

\item {\it Pile-up modelling} (Correlation 0)\\ 
In both ATLAS and CMS, simulated events are reweighted to match the distribution of the average number of interactions per bunch-crossing in data. The corresponding uncertainty is obtained from in situ techniques based on data and simulated event samples. In the ATLAS analyses at \cmenergySeven, the uncertainty due to pile-up is derived from the impact of the reweighting on \myMET. In the ATLAS analyses at \cmenergyEight, this uncertainty is evaluated as a component of the JES, separated into four terms (number of primary vertices, average number of collisions per bunch-crossing, average pile-up energy density in the calorimeter, and \pT\ dependence) since the pile-up calibration (assuming average conditions during 8~\TeV\ data-taking) is applied to both data and simulation before selecting and calibrating the jets~\cite{ATL-PHYS-PUB-2015-049}. 
In CMS, the reweighting uses a model with a free parameter that can be interpreted as an effective cross-section for inelastic $pp$ interactions. This uncertainty is obtained from a fit to the number of additional primary vertices in simulation. In the CMS analyses, this uncertainty is introduced as a nuisance parameter in the fit. The only exception is the \sch\ measurement, where the pile-up uncertainty is estimated from pseudoexperiments. In all cases, the effects of pile-up on the jet energy and the isolation of leptons are taken into account in the jet and lepton uncertainties respectively. The pile-up uncertainty is considered uncorrelated between ATLAS and CMS and between different centre-of-mass energies. It is considered correlated between different production modes~\cite{ATL-PHYS-PUB-2014-020, ATL-PHYS-PUB-2015-049}.

\end{itemize}

\subsection{Systematic uncertainties in theoretical cross-section predictions}
\label{sec:systcat_theo}

The systematic uncertainties in the combined $\absvtb$ value are evaluated from uncertainties in the individual cross-section measurements $\sigmameas$ and the theoretical predictions $\sigmatheo$. The uncertainties associated with $\sigmatheo$ are discussed in Section~\ref{sec:theory}; they are summarised in Table~\ref{tab:xs}. The correlation assumptions for the systematic uncertainties related to the theoretical cross-section are explained below. In Section~\ref{sec:tests}, the stability of the $\absvtb$ combination against variations in the correlations is examined. For clarity, the  correlations are given in parentheses next to the systematic-uncertainty name. These correlations are used in the combination of the $\absvtb$ extractions.\\

\noindent {\bf PDF+$\alpha_{\rm s}$} (Correlation 1 for centre-of-mass energies and 0.5 for production modes) \\ The PDF uncertainty is considered correlated between centre-of-mass energies and 50\% correlated between production modes, since different production modes have one initial-state particle in common (a quark or a gluon), but not both.

\noindent {\bf Renormalisation and factorisation scales} (Correlation 1 for \tch\ and \sch\ and 0 for \mytW) \\ The renormalisation and factorisation scale uncertainties in $\sigmatheo$ are
considered correlated between production modes and centre-of-mass energies, except between the \mytW production mode and the other production modes, where they are considered uncorrelated because the \mytW prediction is computed at a different order in perturbation theory.

\noindent {\bf Top-quark mass} (Correlation 1)\\ 
The uncertainty due to $\topmass$ is evaluated by varying $\topmass$ from its central value of 172.5~\GeV\ by $\pm1$~\GeV\ and evaluating the corresponding change in cross-section using the parameterisation given in Ref.~\cite{Czakon:2013goa}, as discussed in Section~\ref{sec:theory}. This uncertainty is considered correlated between centre-of-mass energies and production modes.

\noindent {\bf $E_{\rm beam}$} (Correlation 1)\\ The uncertainty in the cross-section due to the uncertainty in $E_{\rm beam}$
is estimated by computing the cross-section variation corresponding to a $\pm 1$ standard deviation shift in the beam-energy uncertainty.
It is considered correlated between centre-of-mass energies and production modes.

 \section{Combinations of cross-section measurements}
\label{sec:xscomb}
The cross-section measurements described in
Section~\ref{sec:atlascmsxs} are combined at each centre-of-mass
energy for each production mode. Systematic uncertainties are
categorised and correlation assumptions are employed according to
Section~\ref{sec:systcat}. The combinations are performed using the
iterative BLUE method, as described in Section~\ref{sec:method}. 

As discussed in Section~\ref{sec:method}, the uncertainty in the
measured cross-section associated with the $\topmass$ variation is
not considered in the combination of cross-sections. However, the
shift in the combined cross-section resulting from a variation of $\pm
1~\GeV$ in the top-quark mass is provided where this information is
available. This is calculated by repeating the combination with the
up-shifted and down-shifted input cross-sections. In measurements
where only the magnitude of the shift is available for one experiment,
the sign of the shift is assumed to be the same for both experiments,
as discussed in Section~\ref{sec:systcat_exp}. If the uncertainty associated with the $\topmass$ variation is not
available for one or both of the input measurements, then no shift in the combined cross-section is given.

Additional information about the uncertainties considered in the combination of cross-section measurements is provided in Appendix~\ref{app:allUncs}.

\subsection{Combinations of \tch\ cross-section measurements}
\label{sec:xscomb_tch}
The combination of the ATLAS and CMS \tch\ cross-section measurements
at \cmenergySeven~\cite{TOPQ-2012-21,CMS-TOP-11-021} results, after
one iteration, in 
\begin{eqnarray*}
\finalResulttChanSeven.
\end{eqnarray*}
The relative uncertainty is $\relPrecisiontChanSeven$, which improves
on the uncertainty of 9.1\% in the most precise individual
measurement from CMS~\cite{CMS-TOP-11-021}. The $\chi^2$ for the combination is
$\blueChiSqtChanSeven$, corresponding to a probability of
$\blueChiSqProbtChanSeven\%$. The CMS weight in the combination is
$\CMStChanSevenCoefftChanSeven$, while the ATLAS weight is
$\ATLAStChanSevenCoefftChanSeven$. The overall correlation between the
two measurements is \blueOverallCorrProbtChanSeven\%. The contribution from
each uncertainty category to the total uncertainty in the combined \tch\ cross-section 
measurement at \cmenergySeven\ is shown in Table~\ref{tab:tChanRes}(a).

The combination of the ATLAS and CMS \tch\ cross-section measurements
at \cmenergyEight~\cite{TOPQ-2015-05,CMS-TOP-12-038} results, after
two iterations, in a cross-section of
\begin{eqnarray*}
\finalResulttChanEight.
\end{eqnarray*}
The relative uncertainty is $\relPrecisiontChanEight$, which improves
on the uncertainty of 7.5\% in the most precise individual
measurement from ATLAS~\cite{TOPQ-2015-05}. The $\chi^2$ for the combination is
$\blueChiSqtChanEight$, corresponding to a probability of
$\blueChiSqProbtChanEight\%$. This probability is lower than the
probability of the combination at \cmenergySeven\ because of the differences between the ATLAS and CMS measured cross-sections and their small uncertainties. The ATLAS weight in the combination is
$\ATLAStChanEightCoefftChanEight$, while the CMS weight is
$\CMStChanEightCoefftChanEight$. The overall correlation between the
two measurements is \blueOverallCorrProbtChanEight\%. This is larger
than the correlation between the measurements at \cmenergySeven\ because the statistical and detector uncertainties are lower, thus increasing the importance of the theory modelling uncertainty (which is correlated between the two experiments), as shown in Appendix~\ref{app:tch}. The contribution from each uncertainty category to the
total uncertainty in the combined \tch\ cross-section measurement at 
\cmenergyEight\ is shown in Table~\ref{tab:tChanRes}(b).

\begin{table}[!htbp]
  \caption{Contribution from each uncertainty category to the combined
    \tch\ cross-section ($\sigmatch$) uncertainty at
    (a) \cmenergySeven\ and (b) \cmenergyEight. The total
    uncertainty is computed by adding in quadrature all the individual systematic
    uncertainties (including the uncertainty in the integrated luminosity) and the statistical uncertainty in data. Correlations of systematic uncertainties between
    experiments are presented in Appendix~\ref{app:tch}.}
  \begin{center}
    \begin{minipage}[b]{0.46\hsize}\centering
      \captionof{subfigure}{}
      \begin{tabular}{l | r | r}
\hline
 \hline
 \multicolumn{3}{c}{ $\sigmatch$, $\sqrt{s} = $ 7~\TeV } \\
\hline
\hline
{\bf Combined cross-section} & \multicolumn{2}{c}{67.5~pb} \\
\hline
\hline
\multirow{2}{*}{Uncertainty category} & \multicolumn{2}{c}{Uncertainty} \\\cline{2-3}
 &  ~~[\%] &  [pb] \\
\hline
Data statistical & 3.5 & 2.4\\
\hline
Simulation statistical & 1.4 & 0.9\\
Integrated luminosity & 1.7 & 1.1\\
Theory modelling & 5.1 & 3.5\\
Background normalisation & 1.9 & 1.3\\
Jets & 3.4 & 2.3\\
Detector modelling & 3.4 & 2.3\\
\hline
Total syst.\ unc.\ (excl.\ lumi.) & 7.5 & 5.0\\
Total syst.\ unc.\ (incl.\ lumi.) & 7.6 & 5.2\\
\hline\hline
{\bf Total uncertainty} & 8.4 & 5.7\\
\hline
 \hline
\end{tabular}
     \end{minipage}
    \hfill
    \begin{minipage}[b]{0.46\hsize}\centering
      \captionof{subfigure}{}
      \begin{tabular}{l | r | r}
\hline
 \hline
 \multicolumn{3}{c}{ $\sigmatch$, $\sqrt{s} = $ 8~\TeV } \\
\hline
\hline
{\bf Combined cross-section} & \multicolumn{2}{c}{87.7~pb} \\
\hline
\hline
\multirow{2}{*}{Uncertainty category} & \multicolumn{2}{c}{Uncertainty} \\\cline{2-3}
 &  ~~[\%] &  [pb] \\
\hline
Data statistical & 1.3 & 1.1\\
\hline
Simulation statistical & 0.6 & 0.5\\
Integrated luminosity & 1.7 & 1.5\\
Theory modelling & 5.3 & 4.7\\
Background normalisation & 1.2 & 1.1\\
Jets & 2.6 & 2.3\\
Detector modelling & 1.8 & 1.6\\
\hline
Total syst.\ unc.\ (excl.\ lumi.) & 6.3 & 5.5\\
Total syst.\ unc.\ (incl.\ lumi.) & 6.5 & 5.7\\
\hline\hline
{\bf Total uncertainty} & 6.7 & 5.8\\
\hline
 \hline
\end{tabular}
     \end{minipage}
  \end{center}
  \label{tab:tChanRes}
\end{table}

At both centre-of-mass energies, the uncertainties from theory
modelling are found to be dominant. Details of the central values, the impact of
individual sources of uncertainties, and their correlations between experiments at
\cmenergyComb\ can be found in Appendix~\ref{app:tch}.

The shift in the combined cross-section at \cmenergyEight\ from a variation of $\pm 1~\GeV$ in the top-quark mass is $\mp 0.8$~pb, which is similar to the shifts in the input measurements for the same $\topmass$ variation. The shift in the combined cross-section at \cmenergySeven\ is not evaluated since no estimate is available for the CMS input measurement at \cmenergySeven.

\subsection{Combinations of {$\boldmath \mytW$} cross-section measurements}
\label{sec:xscomb_tW}
The combination of the ATLAS and CMS $\mytW$ cross-section measurements
at \cmenergySeven~\cite{TOPQ-2011-17, CMS-TOP-11-022} yields, after
two iterations, a cross-section of 
\begin{eqnarray*}
\finalResulttWSeven.
\end{eqnarray*}
The relative uncertainty is $\relPrecisiontWSeven$, which improves on the uncertainty of 28\% in the most precise individual
measurement from CMS~\cite{CMS-TOP-11-022}. The $\chi^2$ for the combination is
$\blueChiSqtWSeven$, corresponding to a probability of
$\blueChiSqProbtWSeven\%$. The CMS weight in the combination is
$\CMStWSevenCoefftWSeven$, while the ATLAS weight is
$\ATLAStWSevenCoefftWSeven$. The overall correlation between the
two measurements is \blueOverallCorrProbtWSeven\%. The contribution from
each uncertainty category to the total uncertainty in the combined  \mytW\ 
cross-section measurement at \cmenergySeven\ is shown in
Table~\ref{tab:tWRes}(a).

The combination of the ATLAS and CMS $\mytW$ cross-section measurements
at \cmenergyEight~\cite{TOPQ-2012-20,CMS-TOP-12-040} results, after
two iterations, in
\begin{eqnarray*}
\finalResulttWEight.
\end{eqnarray*}
The relative uncertainty is $\relPrecisiontWEight$, which improves on the uncertainty of 16.5\% in the most precise individual
measurement from ATLAS~\cite{TOPQ-2012-20}. The $\chi^2$ for the combination is
$\blueChiSqtWEight$, corresponding to a probability of
$\blueChiSqProbtWEight\%$. The ATLAS weight in the combination is
$\ATLAStWEightCoefftWEight$, while the CMS weight is
$\CMStWEightCoefftWEight$. The overall correlation between the
two measurements is \blueOverallCorrProbtWEight\%.
Similar to the \tch, this is larger than the correlation between the
measurements at \cmenergySeven\ due to the increased importance of the
theory modelling uncertainties.
The contribution from
each uncertainty category to the total uncertainty in the combined \mytW\ 
cross-section measurement at \cmenergyEight\ is shown in
Table~\ref{tab:tWRes}(b).

\begin{table}[!htbp]
  \caption{Contribution from each uncertainty category to the combined
    \mytW cross-section ($\sigmatW$) uncertainty at
    (a) \cmenergySeven\ and (b) \cmenergyEight. The total
    uncertainty is computed by adding in quadrature all the individual systematic
    uncertainties (including the uncertainty in the integrated luminosity) and the statistical uncertainty in data.
    Correlations of systematic uncertainties between
    experiments are presented in Appendix~\ref{app:tw}.}
  \begin{center}
    \begin{minipage}[b]{0.46\hsize}\centering
      \captionof{subfigure}{}
      \begin{tabular}{l | r | r}
\hline
 \hline
 \multicolumn{3}{c}{ $\sigmatW$, $\sqrt{s} = $ 7~\TeV } \\
\hline
\hline
{\bf Combined cross-section} & \multicolumn{2}{c}{16.3~pb} \\
\hline
\hline
\multirow{2}{*}{Uncertainty category} & \multicolumn{2}{c}{Uncertainty} \\\cline{2-3}
 &  ~~[\%] &  [pb] \\
\hline
Data statistical & 14.0 & 2.3\\
\hline
Simulation statistical & 0.8 & 0.1\\
Integrated luminosity & 4.4 & 0.7\\
Theory modelling & 13.9 & 2.3\\
Background normalisation & 6.0 & 1.0\\
Jets & 11.5 & 1.9\\
Detector modelling & 6.2 & 1.0\\
\hline
Total syst.\ unc.\ (excl.\ lumi.) & 20.0 & 3.3\\
Total syst.\ unc.\ (incl.\ lumi.) & 20.5 & 3.3\\
\hline\hline
{\bf Total uncertainty} & 24.8 & 4.1\\
\hline
 \hline
\end{tabular}
     \end{minipage}
    \hfill
    \begin{minipage}[b]{0.46\hsize}\centering
      \captionof{subfigure}{}
      \begin{tabular}{l | r | r}
\hline
 \hline
 \multicolumn{3}{c}{ $\sigmatW$, $\sqrt{s} = $ 8~\TeV } \\
\hline
\hline
{\bf Combined cross-section} & \multicolumn{2}{c}{23.1~pb} \\
\hline
\hline
\multirow{2}{*}{Uncertainty category} & \multicolumn{2}{c}{Uncertainty} \\\cline{2-3}
 &  ~~[\%] &  [pb] \\
\hline
Data statistical & 4.7 & 1.1\\
\hline
Simulation statistical & 0.8 & 0.2\\
Integrated luminosity & 3.6 & 0.8\\
Theory modelling & 11.8 & 2.7\\
Background normalisation & 2.2 & 0.5\\
Jets & 6.2 & 1.4\\
Detector modelling & 4.9 & 1.1\\
\hline
Total syst.\ unc.\ (excl.\ lumi.) & 14.4 & 3.3\\
Total syst.\ unc.\ (incl.\ lumi.) & 14.8 & 3.4\\
\hline\hline
{\bf Total uncertainty} & 15.6 & 3.6\\
\hline
 \hline
\end{tabular}
     \end{minipage}
  \end{center}
  \label{tab:tWRes}
\end{table}

At both centre-of-mass energies, the uncertainties in the theory
modelling are found to be dominant. The jet uncertainties are also
important. Details of the central values, the impact of
individual sources of uncertainties, and their correlations between experiments at
\cmenergyComb\ are presented in Appendix~\ref{app:tw}.

The shift in the combined cross-section at \cmenergyEight\ from a variation of $\pm
1~\GeV$ in the top-quark mass is $\pm 1.1$~pb, which is similar in magnitude to that in the input measurements for the same $\topmass$ variation. The shift in the combined cross-section at \cmenergySeven\ is not evaluated since no estimates are available for the input measurements at \cmenergySeven.

\subsection{Combination of \sch\ cross-section measurements}
\label{sec:xscomb_sch}

The ATLAS and CMS \sch\ cross-section measurements suffer from large backgrounds, and the cross-section measurements have large uncertainties. Since the systematic uncertainties mainly affect the background prediction, they are not scaled in the iterative BLUE procedure. Only the luminosity uncertainty is scaled with the central value. The combination of the ATLAS and CMS \sch\ cross-section measurements at \cmenergyEight~\cite{TOPQ-2015-01, CMS-TOP-13-009} results, after
two iterations, in a cross-section of

\begin{eqnarray*}
\finalResultsChanEight.
\end{eqnarray*}
The relative uncertainty is $\relPrecisionsChanEight$, very similar to the most precise individual measurement from ATLAS~\cite{TOPQ-2015-01}. The $\chi^2$ for the combination is $\blueChiSqsChanEight$, corresponding to a probability of $\blueChiSqProbsChanEight\%$. 
The ATLAS weight in the combination is $\ATLASsChanEightCoeffsChanEight$, while the CMS weight is $\CMSsChanEightCoeffsChanEight$. The overall correlation between the two measurements is \blueOverallCorrProbsChanEight\%. The contribution from each uncertainty category to the total uncertainty in the combined \sch\ cross-section measurement at \cmenergyEight\ is shown in Table~\ref{tab:sChanRes}.

\begin{table}[!htbp]
  \caption{Contribution from each uncertainty category to the combined
    \sch\ cross-section ($\sigmasch$) uncertainty at \cmenergyEight. The total
    uncertainty is computed by adding in quadrature all the individual systematic
    uncertainties (including the uncertainty in the integrated luminosity) and the statistical uncertainty in data.
    Correlations of systematic uncertainties between
    experiments are presented in Appendix~\ref{app:sch}.}
  \begin{center}
    \begin{tabular}{l | r | r}
\hline
 \hline
 \multicolumn{3}{c}{ $\sigmasch$, $\sqrt{s} = $ 8~\TeV } \\
\hline
\hline
{\bf Combined cross-section} & \multicolumn{2}{c}{4.9~pb} \\
\hline
\hline
\multirow{2}{*}{Uncertainty category} & \multicolumn{2}{c}{Uncertainty} \\\cline{2-3}
 &  ~~[\%] &  [pb] \\
\hline
Data statistical & 16 & 0.8\\
\hline
Simulation statistical & 12 & 0.6\\
Integrated luminosity & 5 & 0.2\\
Theory modelling & 14 & 0.7\\
Background normalisation & 8 & 0.4\\
Jets & 13 & 0.6\\
Detector modelling & 8 & 0.4\\
\hline
Total syst.\ unc.\ (excl.\ lumi.) & 25 & 1.2\\
Total syst.\ unc.\ (incl.\ lumi.) & 25 & 1.2\\
\hline\hline
{\bf Total uncertainty} & 30 & 1.4\\
\hline
 \hline
\end{tabular}
   \end{center}
  \label{tab:sChanRes}
\end{table}

Since the ATLAS measurement has a large weight in the combination, the importance of each uncertainty in the combination is similar to that in the ATLAS measurement, as presented in Appendix~\ref{app:sch}.

The shift in the combined cross-section at \cmenergyEight\ from a variation in the top-quark mass is not evaluated since no estimate is available for the ATLAS input measurement.

\subsection{Summary of cross-section combinations}
\begin{figure}[!ht]
  \begin{center}
    \includegraphics[width=1.0\textwidth]{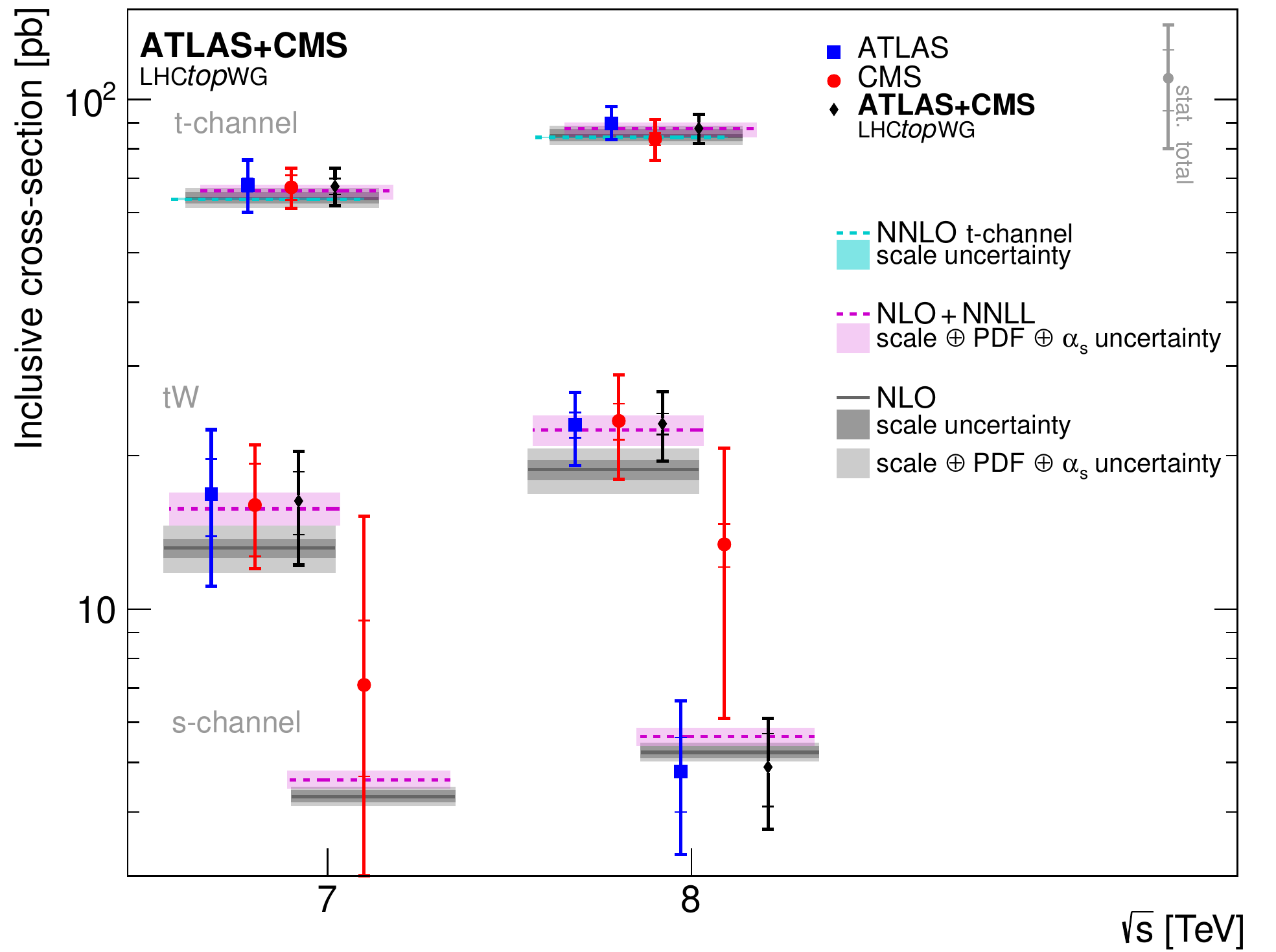}
  \end{center}
  \caption{Single-top-quark cross-section measurements performed by ATLAS and CMS, together with the combined results shown in Sections~\ref{sec:xscomb_tch}$-$\ref{sec:xscomb_sch}. These measurements are compared with the theoretical predictions at NLO and NLO+NNLL for all three production modes and the prediction at NNLO for \tch\ only. The corresponding theoretical uncertainties are also presented. The scale uncertainty for the NNLO prediction is small and is presented as a narrow band under the dashed line.}
  \label{fig:xssum}
\end{figure}

A summary of the cross-sections measured by ATLAS and CMS
and their combinations in all single-top-quark production modes at each
centre-of-mass energy is shown in
Figure~\ref{fig:xssum}. The measurements are compared with the theoretical predictions shown in Table~\ref{tab:xs}: NNLO for \tch\ only, NLO and NLO+NNLL for all three production modes. For the NLO calculation, the renormalisation- and factorisation-scale uncertainties and the sum in quadrature of the contributions from scale, PDF, and $\alpha_{\rm s}$ are shown separately. Only the scale uncertainty is shown for the NNLO calculation. 
For the NLO+NNLL calculation, the sum in quadrature of the contributions from scale, PDF, and $\alpha_{\rm s}$ is shown. All measurements are in good agreement with their corresponding theoretical predictions within their total uncertainties.

The stability of the combinations of the cross-section measurements to variations in
the correlation assumptions, discussed in Section~\ref{sec:systcat}, is  checked for the theory modelling, JES, the most important contributions to the theoretical cross-section predictions
(i.e.\ PDF+$\alpha_{\rm s}$ and scale) and the integrated luminosity. The results
of these tests show that their impacts on the cross-section
combinations are very small, similar to the stability tests
for the combination of the $\absvtb$ values discussed in Section~\ref{sec:tests}.

 \section{Combinations of {\boldmath $\absvtb$} determinations}
\label{sec:vtbcomb}
\newcommand{\expUncvtbtChan}{\ensuremath{0.03}}
\newcommand{\mainresvtbtChan}{\ensuremath{1.02}}
\newcommand{\mainresSquaredvtbtChan}{\ensuremath{1.04}}
\newcommand{\finalResultSquaredvtbtChan}{\ensuremath{|\flv\vtb|^2 = 1.04\pm0.01\;(\text{stat.})\pm0.06\;(\text{syst.})\pm0.01\;(\text{lumi.}) = 1.04\pm0.07}}
\newcommand{\relPrecisionSquaredvtbtChan}{\ensuremath{6.5\%}}
\newcommand{\finalResultvtbtChan}{\ensuremath{|\flv\vtb| = 1.02\pm0.01\;(\text{stat.})\pm0.03\;(\text{syst.})\pm0.01\;(\text{lumi.}) = 1.02\pm0.03}}
\newcommand{\finalResultTotUncvtbtChan}{\ensuremath{1.02\pm0.03}}
\newcommand{\relPrecisionvtbtChan}{\ensuremath{3.2\%}}
\newcommand{\blueChiSqvtbtChan}{\ensuremath{0.63}}
\newcommand{\blueChiSqProbvtbtChan}{\ensuremath{89}}
\newcommand{\ATLAStChanEightCoeffvtbtChan}{\ensuremath{0.09}}
\newcommand{\ATLAStChanEightPullvtbtChan}{\ensuremath{0.25}}
\newcommand{\CMStChanEightCoeffvtbtChan}{\ensuremath{0.48}}
\newcommand{\CMStChanEightPullvtbtChan}{\ensuremath{0.53}}
\newcommand{\ATLAStChanSevenCoeffvtbtChan}{\ensuremath{0.16}}
\newcommand{\ATLAStChanSevenPullvtbtChan}{\ensuremath{0.16}}
\newcommand{\CMStChanSevenCoeffvtbtChan}{\ensuremath{0.27}}
\newcommand{\CMStChanSevenPullvtbtChan}{\ensuremath{-0.80}}
\newcommand{\ATLAStChanEightCMStChanEightRhovtbtChan}{\ensuremath{0.42}}
\newcommand{\ATLAStChanEightATLAStChanSevenRhovtbtChan}{\ensuremath{0.46}}
\newcommand{\ATLAStChanEightCMStChanSevenRhovtbtChan}{\ensuremath{0.40}}
\newcommand{\CMStChanEightATLAStChanSevenRhovtbtChan}{\ensuremath{0.21}}
\newcommand{\CMStChanEightCMStChanSevenRhovtbtChan}{\ensuremath{0.33}}
\newcommand{\ATLAStChanSevenCMStChanSevenRhovtbtChan}{\ensuremath{0.20}}
\newcommand{\relDiffIterativevtbtChan}{\ensuremath{1.56}}
 \newcommand{\expUncvtbtW}{\ensuremath{0.08}}
\newcommand{\mainresvtbtW}{\ensuremath{1.02}}
\newcommand{\mainresSquaredvtbtW}{\ensuremath{1.03}}
\newcommand{\finalResultSquaredvtbtW}{\ensuremath{|\flv\vtb|^2 = 1.03\pm0.05\;(\text{stat.})\pm0.14\;(\text{syst.})\pm0.03\;(\text{lumi.}) = 1.03\pm0.16}}
\newcommand{\relPrecisionSquaredvtbtW}{\ensuremath{15.2\%}}
\newcommand{\finalResultvtbtW}{\ensuremath{|\flv\vtb| = 1.02\pm0.03\;(\text{stat.})\pm0.07\;(\text{syst.})\pm0.02\;(\text{lumi.}) = 1.02\pm0.08}}
\newcommand{\finalResultTotUncvtbtW}{\ensuremath{1.02\pm0.08}}
\newcommand{\relPrecisionvtbtW}{\ensuremath{7.6\%}}
\newcommand{\blueChiSqvtbtW}{\ensuremath{0.02}}
\newcommand{\blueChiSqProbvtbtW}{\ensuremath{100}}
\newcommand{\ATLAStWEightCoeffvtbtW}{\ensuremath{0.08}}
\newcommand{\ATLAStWEightPullvtbtW}{\ensuremath{0.11}}
\newcommand{\CMStWEightCoeffvtbtW}{\ensuremath{0.61}}
\newcommand{\CMStWEightPullvtbtW}{\ensuremath{-0.05}}
\newcommand{\ATLAStWSevenCoeffvtbtW}{\ensuremath{0.16}}
\newcommand{\ATLAStWSevenPullvtbtW}{\ensuremath{-0.06}}
\newcommand{\CMStWSevenCoeffvtbtW}{\ensuremath{0.15}}
\newcommand{\CMStWSevenPullvtbtW}{\ensuremath{0.08}}
\newcommand{\ATLAStWEightCMStWEightRhovtbtW}{\ensuremath{0.43}}
\newcommand{\ATLAStWEightATLAStWSevenRhovtbtW}{\ensuremath{0.15}}
\newcommand{\ATLAStWEightCMStWSevenRhovtbtW}{\ensuremath{0.12}}
\newcommand{\CMStWEightATLAStWSevenRhovtbtW}{\ensuremath{0.35}}
\newcommand{\CMStWEightCMStWSevenRhovtbtW}{\ensuremath{0.29}}
\newcommand{\ATLAStWSevenCMStWSevenRhovtbtW}{\ensuremath{0.17}}
\newcommand{\relDiffIterativevtbtW}{\ensuremath{1.55}}
 \newcommand{\expUncvtbsChan}{\ensuremath{0.14}}
\newcommand{\mainresvtbsChan}{\ensuremath{0.97}}
\newcommand{\mainresSquaredvtbsChan}{\ensuremath{0.93}}
\newcommand{\finalResultSquaredvtbsChan}{\ensuremath{|\flv\vtb|^2 = 0.93\pm0.15\;(\text{stat.})\pm0.23\;(\text{syst.})\pm0.05\;(\text{lumi.}) = 0.93\pm0.28}}
\newcommand{\relPrecisionSquaredvtbsChan}{\ensuremath{29.7\%}}
\newcommand{\finalResultvtbsChan}{\ensuremath{|\flv\vtb| = 0.97\pm0.08\;(\text{stat.})\pm0.12\;(\text{syst.})\pm0.02\;(\text{lumi.}) = 0.97\pm0.14}}
\newcommand{\finalResultTotUncvtbsChan}{\ensuremath{0.97\pm0.14}}
\newcommand{\relPrecisionvtbsChan}{\ensuremath{14.8\%}}
\newcommand{\blueChiSqvtbsChan}{\ensuremath{1.42}}
\newcommand{\blueChiSqProbvtbsChan}{\ensuremath{23}}
\newcommand{\ATLASsChanEightCoeffvtbsChan}{\ensuremath{0.99}}
\newcommand{\ATLASsChanEightPullvtbsChan}{\ensuremath{-1.19}}
\newcommand{\CMSsChanEightCoeffvtbsChan}{\ensuremath{0.01}}
\newcommand{\CMSsChanEightPullvtbsChan}{\ensuremath{1.19}}
\newcommand{\ATLASsChanEightCMSsChanEightRhovtbsChan}{\ensuremath{0.15}}
\newcommand{\relDiffIterativevtbsChan}{\ensuremath{3.60}}
 \newcommand{\finalResultSquaredvtbtChanTheo}{\ensuremath{|\flv\vtb|^2 = 1.04\pm0.03\;(\text{stat.})\pm0.04\;(\text{syst.})\pm0.00\;(\text{lumi.}) = 1.04\pm0.05}}
\newcommand{\mainresSquaredvtbtChanTheo}{\ensuremath{1.04}}
\newcommand{\UncStatSquaredvtbtChanTheo}{\ensuremath{0.03}}
\newcommand{\UncSystSquaredvtbtChanTheo}{\ensuremath{0.04}}
\newcommand{\UncTotalSquaredvtbtChanTheo}{\ensuremath{0.05}}
\newcommand{\theoUncvtbtChanTheo}{\ensuremath{0.02}}
\newcommand{\finalResultvtbtChanTheo}{\ensuremath{|\flv\vtb| = 1.02\pm0.01\;(\text{stat.})\pm0.02\;(\text{syst.})\pm0.00\;(\text{lumi.}) = 1.02\pm0.02}}
\newcommand{\finalResultTotUncvtbtChanTheo}{\ensuremath{1.02\pm0.02}}
\newcommand{\relPrecisionvtbtChanTheo}{\ensuremath{2.4\%}}
\newcommand{\blueChiSqvtbtChanTheo}{\ensuremath{1.54}}
\newcommand{\blueChiSqProbvtbtChanTheo}{\ensuremath{67}}
\newcommand{\ATLAStChanEightCoeffvtbtChanTheo}{\ensuremath{0.23}}
\newcommand{\ATLAStChanEightPullvtbtChanTheo}{\ensuremath{0.58}}
\newcommand{\CMStChanEightCoeffvtbtChanTheo}{\ensuremath{0.26}}
\newcommand{\CMStChanEightPullvtbtChanTheo}{\ensuremath{0.46}}
\newcommand{\ATLAStChanSevenCoeffvtbtChanTheo}{\ensuremath{0.23}}
\newcommand{\ATLAStChanSevenPullvtbtChanTheo}{\ensuremath{0.31}}
\newcommand{\CMStChanSevenCoeffvtbtChanTheo}{\ensuremath{0.28}}
\newcommand{\CMStChanSevenPullvtbtChanTheo}{\ensuremath{-1.23}}
\newcommand{\ATLAStChanEightCMStChanEightRhovtbtChanTheo}{\ensuremath{0.43}}
\newcommand{\ATLAStChanEightATLAStChanSevenRhovtbtChanTheo}{\ensuremath{0.44}}
\newcommand{\ATLAStChanEightCMStChanSevenRhovtbtChanTheo}{\ensuremath{0.44}}
\newcommand{\CMStChanEightATLAStChanSevenRhovtbtChanTheo}{\ensuremath{0.44}}
\newcommand{\CMStChanEightCMStChanSevenRhovtbtChanTheo}{\ensuremath{0.44}}
\newcommand{\ATLAStChanSevenCMStChanSevenRhovtbtChanTheo}{\ensuremath{0.45}}
\newcommand{\relDiffIterativevtbtChanTheo}{\ensuremath{1.78}}
 \newcommand{\finalResultSquaredvtbtWTheo}{\ensuremath{|\flv\vtb|^2 = 1.04\pm0.03\;(\text{stat.})\pm0.07\;(\text{syst.})\pm0.00\;(\text{lumi.}) = 1.04\pm0.08}}
\newcommand{\mainresSquaredvtbtWTheo}{\ensuremath{1.04}}
\newcommand{\UncStatSquaredvtbtWTheo}{\ensuremath{0.03}}
\newcommand{\UncSystSquaredvtbtWTheo}{\ensuremath{0.07}}
\newcommand{\UncTotalSquaredvtbtWTheo}{\ensuremath{0.08}}
\newcommand{\theoUncvtbtWTheo}{\ensuremath{0.04}}
\newcommand{\finalResultvtbtWTheo}{\ensuremath{|\flv\vtb| = 1.02\pm0.01\;(\text{stat.})\pm0.04\;(\text{syst.})\pm0.00\;(\text{lumi.}) = 1.02\pm0.04}}
\newcommand{\finalResultTotUncvtbtWTheo}{\ensuremath{1.02\pm0.04}}
\newcommand{\relPrecisionvtbtWTheo}{\ensuremath{3.8\%}}
\newcommand{\blueChiSqvtbtWTheo}{\ensuremath{0.57}}
\newcommand{\blueChiSqProbvtbtWTheo}{\ensuremath{90}}
\newcommand{\ATLAStWEightCoeffvtbtWTheo}{\ensuremath{0.08}}
\newcommand{\ATLAStWEightPullvtbtWTheo}{\ensuremath{0.58}}
\newcommand{\CMStWEightCoeffvtbtWTheo}{\ensuremath{0.38}}
\newcommand{\CMStWEightPullvtbtWTheo}{\ensuremath{-0.20}}
\newcommand{\ATLAStWSevenCoeffvtbtWTheo}{\ensuremath{0.18}}
\newcommand{\ATLAStWSevenPullvtbtWTheo}{\ensuremath{-0.38}}
\newcommand{\CMStWSevenCoeffvtbtWTheo}{\ensuremath{0.36}}
\newcommand{\CMStWSevenPullvtbtWTheo}{\ensuremath{0.27}}
\newcommand{\ATLAStWEightCMStWEightRhovtbtWTheo}{\ensuremath{0.68}}
\newcommand{\ATLAStWEightATLAStWSevenRhovtbtWTheo}{\ensuremath{0.70}}
\newcommand{\ATLAStWEightCMStWSevenRhovtbtWTheo}{\ensuremath{0.69}}
\newcommand{\CMStWEightATLAStWSevenRhovtbtWTheo}{\ensuremath{0.70}}
\newcommand{\CMStWEightCMStWSevenRhovtbtWTheo}{\ensuremath{0.70}}
\newcommand{\ATLAStWSevenCMStWSevenRhovtbtWTheo}{\ensuremath{0.72}}
\newcommand{\relDiffIterativevtbtWTheo}{\ensuremath{1.70}}
 \newcommand{\finalResultSquaredvtbsChanTheo}{\ensuremath{|\flv\vtb|^2 = 0.97\pm0.05\;(\text{stat.})\pm0.04\;(\text{syst.})\pm0.00\;(\text{lumi.}) = 0.97\pm0.07}}
\newcommand{\mainresSquaredvtbsChanTheo}{\ensuremath{0.97}}
\newcommand{\UncStatSquaredvtbsChanTheo}{\ensuremath{0.05}}
\newcommand{\UncSystSquaredvtbsChanTheo}{\ensuremath{0.04}}
\newcommand{\UncTotalSquaredvtbsChanTheo}{\ensuremath{0.07}}
\newcommand{\theoUncvtbsChanTheo}{\ensuremath{0.03}}
\newcommand{\finalResultvtbsChanTheo}{\ensuremath{|\flv\vtb| = 0.98\pm0.02\;(\text{stat.})\pm0.02\;(\text{syst.})\pm0.00\;(\text{lumi.}) = 0.98\pm0.03}}
\newcommand{\finalResultTotUncvtbsChanTheo}{\ensuremath{0.98\pm0.03}}
\newcommand{\relPrecisionvtbsChanTheo}{\ensuremath{3.4\%}}
\newcommand{\blueChiSqvtbsChanTheo}{\ensuremath{300.77}}
\newcommand{\blueChiSqProbvtbsChanTheo}{\ensuremath{0}}
\newcommand{\ATLASsChanEightCoeffvtbsChanTheo}{\ensuremath{0.97}}
\newcommand{\ATLASsChanEightPullvtbsChanTheo}{\ensuremath{-17.34}}
\newcommand{\CMSsChanEightCoeffvtbsChanTheo}{\ensuremath{0.03}}
\newcommand{\CMSsChanEightPullvtbsChanTheo}{\ensuremath{17.34}}
\newcommand{\ATLASsChanEightCMSsChanEightRhovtbsChanTheo}{\ensuremath{0.55}}
\newcommand{\relDiffIterativevtbsChanTheo}{\ensuremath{18.56}}
 \newcommand{\expUncvtbCombtChan}{\ensuremath{0.04}}
\newcommand{\theoryUncvtbCombtChan}{\ensuremath{0.02}}
\newcommand{\totUncvtbCombtChan}{\ensuremath{0.04}}
\newcommand{\mainresvtbCombtChan}{\ensuremath{1.02}}
\newcommand{\mainresSquaredvtbCombtChan}{\ensuremath{1.04}}
\newcommand{\finalResultSquaredvtbCombtChan}{\ensuremath{|\flv\vtb|^2 = 1.04\pm0.01\;(\text{stat.})\pm0.08\;(\text{syst.})\pm0.01\;(\text{lumi.}) = 1.04\pm0.08}}
\newcommand{\finalResultSquaredWithTheoryXSvtbCombtChan}{\ensuremath{|\flv\vtb|^2 = 1.04\pm0.01\;(\text{stat.})\pm0.06\;(\text{syst.})\pm0.01\;(\text{lumi.})\pm0.04\;(\text{theo.}) = 1.04\pm0.08}}
\newcommand{\relPrecisionSquaredForTheoryXSvtbCombtChan}{\ensuremath{4.2\%}}
\newcommand{\relPrecisionForTheoryXSvtbCombtChan}{\ensuremath{2.1\%}}
\newcommand{\relPrecisionSquaredvtbCombtChan}{\ensuremath{7.7\%}}
\newcommand{\finalResultvtbCombtChan}{\ensuremath{|\flv\vtb| = 1.02\pm0.01\;(\text{stat.})\pm0.04\;(\text{syst.})\pm0.01\;(\text{lumi.}) = 1.02\pm0.04}}
\newcommand{\finalResultWithTheoryXSvtbCombtChan}{\ensuremath{|\flv\vtb|=~&~1.02\pm0.01\;(\text{stat.})\pm0.03\;(\text{syst.})\pm0.01\;(\text{lumi.})\pm0.02\;(\text{theo.}) \\=~&~1.02\pm0.04\;(\text{meas.})\pm0.02\;(\text{theo.}) = 1.02\pm0.04}}
\newcommand{\finalResultWithTheoryXSCompactvtbCombtChan}{\ensuremath{|\flv\vtb| = 1.02\pm0.04\;(\text{meas.})\pm0.02\;(\text{theo.})}}
\newcommand{\finalResultTotUncvtbCombtChan}{\ensuremath{1.02\pm0.04}}
\newcommand{\relPrecisionvtbCombtChan}{\ensuremath{3.9\%}}
\newcommand{\blueChiSqvtbCombtChan}{\ensuremath{0.63}}
\newcommand{\blueChiSqProbvtbCombtChan}{\ensuremath{89}}
\newcommand{\ATLAStChanEightCoeffvtbCombtChan}{\ensuremath{0.09}}
\newcommand{\ATLAStChanEightPullvtbCombtChan}{\ensuremath{0.25}}
\newcommand{\CMStChanEightCoeffvtbCombtChan}{\ensuremath{0.49}}
\newcommand{\CMStChanEightPullvtbCombtChan}{\ensuremath{0.54}}
\newcommand{\ATLAStChanSevenCoeffvtbCombtChan}{\ensuremath{0.15}}
\newcommand{\ATLAStChanSevenPullvtbCombtChan}{\ensuremath{0.16}}
\newcommand{\CMStChanSevenCoeffvtbCombtChan}{\ensuremath{0.27}}
\newcommand{\CMStChanSevenPullvtbCombtChan}{\ensuremath{-0.80}}
\newcommand{\ATLAStChanEightCMStChanEightRhovtbCombtChan}{\ensuremath{0.54}}
\newcommand{\ATLAStChanEightATLAStChanSevenRhovtbCombtChan}{\ensuremath{0.55}}
\newcommand{\ATLAStChanEightCMStChanSevenRhovtbCombtChan}{\ensuremath{0.52}}
\newcommand{\CMStChanEightATLAStChanSevenRhovtbCombtChan}{\ensuremath{0.33}}
\newcommand{\CMStChanEightCMStChanSevenRhovtbCombtChan}{\ensuremath{0.44}}
\newcommand{\ATLAStChanSevenCMStChanSevenRhovtbCombtChan}{\ensuremath{0.31}}
\newcommand{\relDiffIterativevtbCombtChan}{\ensuremath{1.45}}
 \newcommand{\expUncvtbCombtW}{\ensuremath{0.08}}
\newcommand{\theoryUncvtbCombtW}{\ensuremath{0.04}}
\newcommand{\totUncvtbCombtW}{\ensuremath{0.09}}
\newcommand{\mainresvtbCombtW}{\ensuremath{1.02}}
\newcommand{\mainresSquaredvtbCombtW}{\ensuremath{1.03}}
\newcommand{\finalResultSquaredvtbCombtW}{\ensuremath{|\flv\vtb|^2 = 1.03\pm0.05\;(\text{stat.})\pm0.16\;(\text{syst.})\pm0.03\;(\text{lumi.}) = 1.03\pm0.17}}
\newcommand{\finalResultSquaredWithTheoryXSvtbCombtW}{\ensuremath{|\flv\vtb|^2 = 1.03\pm0.05\;(\text{stat.})\pm0.14\;(\text{syst.})\pm0.03\;(\text{lumi.})\pm0.07\;(\text{theo.}) = 1.03\pm0.17}}
\newcommand{\relPrecisionSquaredForTheoryXSvtbCombtW}{\ensuremath{7.2\%}}
\newcommand{\relPrecisionForTheoryXSvtbCombtW}{\ensuremath{3.6\%}}
\newcommand{\relPrecisionSquaredvtbCombtW}{\ensuremath{16.8\%}}
\newcommand{\finalResultvtbCombtW}{\ensuremath{|\flv\vtb| = 1.02\pm0.03\;(\text{stat.})\pm0.08\;(\text{syst.})\pm0.02\;(\text{lumi.}) = 1.02\pm0.09}}
\newcommand{\finalResultWithTheoryXSvtbCombtW}{\ensuremath{|\flv\vtb|=~&~1.02\pm0.03\;(\text{stat.})\pm0.07\;(\text{syst.})\pm0.02\;(\text{lumi.})\pm0.04\;(\text{theo.}) \\=~&~1.02\pm0.09\;(\text{meas.})\pm0.04\;(\text{theo.}) = 1.02\pm0.09}}
\newcommand{\finalResultWithTheoryXSCompactvtbCombtW}{\ensuremath{|\flv\vtb| = 1.02\pm0.09\;(\text{meas.})\pm0.04\;(\text{theo.})}}
\newcommand{\finalResultTotUncvtbCombtW}{\ensuremath{1.02\pm0.09}}
\newcommand{\relPrecisionvtbCombtW}{\ensuremath{8.4\%}}
\newcommand{\blueChiSqvtbCombtW}{\ensuremath{0.02}}
\newcommand{\blueChiSqProbvtbCombtW}{\ensuremath{100}}
\newcommand{\ATLAStWEightCoeffvtbCombtW}{\ensuremath{0.08}}
\newcommand{\ATLAStWEightPullvtbCombtW}{\ensuremath{0.11}}
\newcommand{\CMStWEightCoeffvtbCombtW}{\ensuremath{0.61}}
\newcommand{\CMStWEightPullvtbCombtW}{\ensuremath{-0.05}}
\newcommand{\ATLAStWSevenCoeffvtbCombtW}{\ensuremath{0.15}}
\newcommand{\ATLAStWSevenPullvtbCombtW}{\ensuremath{-0.05}}
\newcommand{\CMStWSevenCoeffvtbCombtW}{\ensuremath{0.16}}
\newcommand{\CMStWSevenPullvtbCombtW}{\ensuremath{0.08}}
\newcommand{\ATLAStWEightCMStWEightRhovtbCombtW}{\ensuremath{0.49}}
\newcommand{\ATLAStWEightATLAStWSevenRhovtbCombtW}{\ensuremath{0.22}}
\newcommand{\ATLAStWEightCMStWSevenRhovtbCombtW}{\ensuremath{0.20}}
\newcommand{\CMStWEightATLAStWSevenRhovtbCombtW}{\ensuremath{0.39}}
\newcommand{\CMStWEightCMStWSevenRhovtbCombtW}{\ensuremath{0.34}}
\newcommand{\ATLAStWSevenCMStWSevenRhovtbCombtW}{\ensuremath{0.21}}
\newcommand{\relDiffIterativevtbCombtW}{\ensuremath{1.54}}
 \newcommand{\expUncvtbCombsChan}{\ensuremath{0.12}}
\newcommand{\theoryUncvtbCombsChan}{\ensuremath{0.02}}
\newcommand{\totUncvtbCombsChan}{\ensuremath{0.15}}
\newcommand{\mainresvtbCombsChan}{\ensuremath{0.97}}
\newcommand{\mainresSquaredvtbCombsChan}{\ensuremath{0.93}}
\newcommand{\finalResultSquaredvtbCombsChan}{\ensuremath{|\flv\vtb|^2 = 0.93\pm0.15\;(\text{stat.})\pm0.23\;(\text{syst.})\pm0.05\;(\text{lumi.}) = 0.93\pm0.28}}
\newcommand{\finalResultSquaredWithTheoryXSvtbCombsChan}{\ensuremath{|\flv\vtb|^2 = 0.93\pm0.15\;(\text{stat.})\pm0.23\;(\text{syst.})\pm0.05\;(\text{lumi.})\pm0.04\;(\text{theo.}) = 0.93\pm0.28}}
\newcommand{\relPrecisionSquaredForTheoryXSvtbCombsChan}{\ensuremath{4.6\%}}
\newcommand{\relPrecisionForTheoryXSvtbCombsChan}{\ensuremath{2.3\%}}
\newcommand{\relPrecisionSquaredvtbCombsChan}{\ensuremath{30.1\%}}
\newcommand{\finalResultvtbCombsChan}{\ensuremath{|\flv\vtb| = 0.97\pm0.08\;(\text{stat.})\pm0.12\;(\text{syst.})\pm0.02\;(\text{lumi.}) = 0.97\pm0.15}}
\newcommand{\finalResultWithTheoryXSvtbCombsChan}{\ensuremath{|\flv\vtb|=~&~0.97\pm0.08\;(\text{stat.})\pm0.12\;(\text{syst.})\pm0.02\;(\text{lumi.})\pm0.02\;(\text{theo.}) \\=~&~0.97\pm0.15\;(\text{meas.})\pm0.02\;(\text{theo.}) = 0.97\pm0.15}}
\newcommand{\finalResultWithTheoryXSCompactvtbCombsChan}{\ensuremath{|\flv\vtb| = 0.97\pm0.15\;(\text{meas.})\pm0.02\;(\text{theo.})}}
\newcommand{\finalResultTotUncvtbCombsChan}{\ensuremath{0.97\pm0.15}}
\newcommand{\relPrecisionvtbCombsChan}{\ensuremath{15.0\%}}
\newcommand{\blueChiSqvtbCombsChan}{\ensuremath{1.42}}
\newcommand{\blueChiSqProbvtbCombsChan}{\ensuremath{23}}
\newcommand{\ATLASsChanEightCoeffvtbCombsChan}{\ensuremath{0.99}}
\newcommand{\ATLASsChanEightPullvtbCombsChan}{\ensuremath{-1.19}}
\newcommand{\CMSsChanEightCoeffvtbCombsChan}{\ensuremath{0.01}}
\newcommand{\CMSsChanEightPullvtbCombsChan}{\ensuremath{1.19}}
\newcommand{\ATLASsChanEightCMSsChanEightRhovtbCombsChan}{\ensuremath{0.15}}
\newcommand{\relDiffIterativevtbCombsChan}{\ensuremath{3.88}}
 The measured cross-section for a given single-top-quark production mode, $\sigma_{\rm meas.}$, has
a linear dependence on $\vtbsq$ as defined in Eq.~(\ref{eq:vtb}). Thus, a value of $\vtbsq$ is extracted from each cross-section measurement and the corresponding theoretical prediction (presented in
Sections~\ref{sec:atlascmsxs} and \ref{sec:theory}
respectively). These values are then combined per channel, and in an
overall $\vtbsq$ combination. In the overall combination, the value from the CMS measurement of $\sigmasch$ is excluded. The reason for excluding the CMS \sch\ analysis from the overall $\vtbsq$ combination is that, at the same centre-of-mass energy, the CMS \tch\ determination has strong correlations with the \sch\ determination, which contains relatively large uncertainties. The strong
correlation between these two measurements makes the combined $\vtbsq$ value strongly
dependent on the correlation assumptions for the dominant uncertainties.
This results in a large variation of the combined $\vtbsq$ value for different correlation assumptions.

All uncertainties in  $\sigmameas$ and $\sigmatheo$ are propagated to the $\vtbsq$ values,
taking into account the correlations described
in Section~\ref{sec:systcat}. The combined value of $\vtbsq$ is evaluated using the reference theoretical
cross-section central values marked with a $^\dagger$ in
Table~\ref{tab:xs}, where it can also be seen that the $E_{\rm
  beam}$ uncertainty is negligible compared to other
uncertainties. For the most precise measurements (i.e.\ for
$\sigmatch$ cross-section measurements at \cmenergyEight), which have a large expected impact on the combination,
the other theoretical calculations from Table~\ref{tab:xs} are used as
cross-checks.

Table~\ref{tab:uncVtb} contains a summary of the individual $\vtbsq$
determinations that are the inputs to the overall $\vtbsq$ combination, together with their experimental and theoretical uncertainties using the reference theoretical cross-sections and uncertainties. For the same processes and at the same centre-of-mass
energies, there are some important differences between uncertainty
categories. In analyses based on \tch\ events at \cmenergySeven, the data
statistical uncertainty is larger in CMS than in ATLAS because the two experiments use data samples of different size. Differences in the category of jet uncertainties are due to the evaluation of the JES uncertainty in ATLAS using pseudoexperiments, while this uncertainty is introduced as a nuisance parameter in the fit in CMS. At \cmenergyEight, the difference between ATLAS and CMS in the background-normalisation category is due to the different techniques used to estimate each background
uncertainty. Additional details are discussed in Appendix~\ref{app:tch}. In the CMS
\mytW\ analysis at \cmenergySeven, the uncertainty
associated with the size of the simulated samples is evaluated as part
of the total statistical uncertainty. The large difference in the
pile-up uncertainty between ATLAS and CMS is due to the different
methods used to assess this uncertainty, as discussed in
Section~\ref{sec:systcat_exp}.
At \cmenergyEight, the sizes of the data and simulated samples used in the CMS \mytW\ analysis are smaller than in the ATLAS analysis, resulting in larger data and simulation statistical uncertainties. The large difference between the two experiments in the category of jet uncertainties arises because the JES uncertainty in ATLAS is evaluated in different categories mostly using pseudoexperiments, while in CMS the JES uncertainty is introduced as a nuisance parameter in the fit. Further details are discussed in Appendix~\ref{app:tw}.
In the CMS \sch\ analysis, the uncertainty associated with the size of the simulated samples is evaluated as part of the total statistical uncertainty. More details are discussed in Appendix~\ref{app:sch}.

\begin{sidewaystable}[!htbp]
  \centering
  \caption{Results of the ATLAS and CMS individual $\vtbsq$ determinations that are the inputs to the overall $\vtbsq$
     combination together with their experimental uncertainties. The values  of $\vtbsq$ may
    slightly differ from those published for the different analyses since in this paper the theoretical cross-sections used are those marked with $^\dagger$ in Table~\ref{tab:xs}. Experimental uncertainties contributing
    less than 1\% are denoted by $<$0.01. Entries with $-$ mean
    that this uncertainty was not evaluated for this
    analysis. Descriptions of the background categories and of the
    correlations of systematic uncertainties between experiments are presented in
    Appendix~\ref{app:allUncs}.}
  \IfPackagesLoaded{adjustbox}{\begin{adjustbox}{max width=0.80\textwidth}}{}
    \begin{tabular}{l|rrrrrrrrr}
\hline
\hline
& $t$-channel& $t$-channel& $t$-channel& $t$-channel& $tW$& $tW$& $tW$& $tW$& $s$-channel\\
& ATLAS& CMS& ATLAS& CMS& ATLAS& CMS& ATLAS& CMS& ATLAS\\
& 8 TeV& 8 TeV& 7 TeV& 7 TeV& 8 TeV& 8 TeV& 7 TeV& 7 TeV& 8 TeV\\
\hline\hline
\T\B{\boldmath $\vtbsq$}&1.06&0.99&1.06&1.05&1.03&1.05&1.07&1.02&0.92\\
\hline
\hline
\textbf{Uncertainties:}  & & & & & & & & &\\
\hline
~~\textbf{Data statistical}& 0.01& 0.03& 0.03& 0.06& 0.06& 0.09& 0.18& 0.21& 0.15\\
\hline
~~\textbf{Simulation statistical}& 0.01& 0.01& 0.02& 0.02& 0.01& 0.03& 0.02& $-$& 0.11\\
\hline
~~\textbf{Integrated luminosity}& 0.02& 0.03& 0.02& 0.02& 0.05& 0.03& 0.07& 0.04& 0.05\\
\hline
~~\textbf{Theory modelling}& & & & & & & & & \\
~~~~ISR/FSR, ren./fact.\ scale& 0.04& 0.02& 0.03& 0.04& 0.09& 0.13& 0.05& 0.03& 0.06\\
~~~~NLO match., generator& 0.03& 0.05& 0.02& 0.04& 0.03& $-$& 0.11& $-$& 0.10\\
~~~~Parton shower& 0.02& $-$& $-$& 0.01& 0.02& 0.15& 0.16& 0.10& 0.02\\
~~~~PDF& 0.01& 0.02& 0.03& 0.01& 0.01& 0.02& 0.02& 0.02& 0.03\\
~~~~DS/DR scheme& $-$& $-$& $-$& $-$& 0.04& 0.02& $-$& 0.06& $-$\\
~~~~Top-quark $p_{\rm T}$ rew.& $-$& $-$& $-$& $-$& $-$& $<$0.01& $-$& $-$& $-$\\
\hline
~~\textbf{Background normalisation}& & & & & & & & & \\
~~~~Top-quark bkg.& $<$0.01& 0.02& 0.02& 0.01& 0.02& 0.02& 0.06& 0.06& 0.05\\
~~~~Other bkg.\ from sim.& 0.01& $<$0.01& $<$0.01& 0.03& 0.02& 0.03& 0.09& 0.04& 0.05\\
~~~~Bkg.\ from data& $<$0.01& 0.02& 0.01& 0.01& $<$0.01& $-$& 0.02& $-$& 0.01\\
\hline
~~\textbf{Jets}& & & & & & & & & \\
~~~~JES common& 0.03& 0.04& 0.08& 0.01& 0.05& 0.04& 0.17& 0.15& 0.05\\
~~~~JES flavour& $<$0.01& $-$& 0.02& $-$& 0.02& $-$& $-$& $-$& 0.01\\
~~~~JetID& $<$0.01& $-$& 0.01& $-$& $<$0.01& $-$& 0.05& $-$& 0.01\\
~~~~JER& $<$0.01& 0.01& 0.02& $<$0.01& 0.07& 0.01& 0.02& 0.04& 0.11\\
\hline
~~\textbf{Detector modelling}& & & & & & & & & \\
~~~~Leptons& 0.02& 0.01& 0.03& 0.04& 0.03& 0.02& 0.07& 0.05& 0.02\\
~~~~HLT (had.\ part)& $-$& $-$& $-$& 0.02& $-$& $-$& $-$& $-$& $-$\\
~~~~\myMET scale& $<$0.01& $<$0.01& 0.03& $<$0.01& 0.06& $<$0.01& $-$& 0.03& 0.01\\
~~~~\myMET res.& $<$0.01& $-$& $-$& $-$& $<$0.01& $-$& $-$& $-$& 0.01\\
~~~~\btag& 0.01& 0.02& 0.04& 0.02& 0.01& 0.01& $-$& 0.02& 0.07\\
~~~~Pile-up& $<$0.01& 0.01& $<$0.01& 0.01& 0.03& $<$0.01& 0.11& 0.01& 0.01\\
\hline
~~\textbf{Top-quark mass}& 0.01& $<$0.01& 0.01& $-$& 0.05& 0.05& $-$& $-$& $-$\\
\hline
~~\textbf{Theoretical cross-section}& & & & & & & & & \\
~~~~PDF+$\alpha_{\rm s}$& 0.03& 0.03& 0.04& 0.04& 0.06& 0.07& 0.08& 0.07& 0.03\\
~~~~Ren./fact.\ scale& 0.03& 0.03& 0.03& 0.03& 0.03& 0.03& 0.03& 0.03& 0.02\\
~~~~Top-quark mass& 0.01& 0.01& 0.01& 0.01& 0.02& 0.02& 0.02& 0.02& 0.02\\
~~~~$E_{\text{beam}}$& $<$0.01& $<$0.01& $<$0.01& $<$0.01& $<$0.01& $<$0.01& $<$0.01& $<$0.01& $<$0.01\\
\hline
\hline
\textbf{Total systematic uncertainty}&0.09&0.09&0.13&0.10&0.18&0.23&0.34&0.24&0.24\\
\hline
\textbf{Total uncertainty}&0.09&0.10&0.13&0.12&0.19&0.24&0.38&0.32&0.28\\
\hline
\hline
\end{tabular}
     \IfPackagesLoaded{adjustbox}{\end{adjustbox}}{}
  \label{tab:uncVtb}
\end{sidewaystable}

\subsection{Results} 
\label{sec:vtbresult}
The combination of $\vtbsq$ is performed using the inputs from
all three single-top-quark production modes. Using the same method, the combination of $\vtbsq$ is also performed separately for each production mode for comparison.

Combining the $\vtbsq$ values extracted from the \tch\ and \mytW\ cross-section measurements at \cmenergyComb\ from ATLAS and CMS, as well as the ATLAS \sch\ measurement at \cmenergyEight, results in
\begin{linenomath}
  \begin{equation*}
    \finalResultSquaredWithTheoryXSvtbComb,
    \label{eq:resultSqd}
  \end{equation*}
\end{linenomath}
with a relative uncertainty of $\relPrecisionSquaredvtbComb$. The contribution from each experimental uncertainty category to the
total uncertainty in the combined $\vtbsq$ value is shown in Table~\ref{tab:vtbRes}. The
theory modelling uncertainties in signal and background processes, discussed in
Section~\ref{sec:systcat_exp}, dominate the experimental
uncertainty and the total uncertainty. The theoretical cross-section uncertainty is the
second-largest contribution to the total uncertainty in the combined $\vtbsq$ value. Changes in the combined $\vtbsq$ value from using alternative NNLO and NLO+NNLL theoretical predictions for the \tch\ are less than 1\%.

\begin{table}[htbp]
  \begin{center}
    \caption{Contributions from each experimental and theoretical
      uncertainty category to the overall $\vtbsq$ combination. The total uncertainty is
      computed by adding in quadrature all of the individual systematic
      uncertainties (including the integrated luminosity and theoretical
      cross-section) and the statistical uncertainty in data.}
    \begin{tabular}{l | r | r}
\hline
 \hline
\T\B
{\bf Combined} {\boldmath $\vtbsq$} & \multicolumn{2}{c}{1.05} \\
\hline
\hline
\multirow{2}{*}{Uncertainty category} & \multicolumn{2}{c}{Uncertainty} \\\cline{2-3}
 &  ~~[\%] &  $\Delta\vtbsq$ \\
\hline
Data statistical & 1.8 & 0.02\\
\hline
Simulation statistical & 0.9 & 0.01\\
Integrated luminosity & 1.3 & 0.01\\
Theory modelling & 4.5 & 0.05\\
Background normalisation & 1.3 & 0.01\\
Jets & 2.6 & 0.03\\
Detector modelling & 1.6 & 0.02\\
Top-quark mass & 0.7 & 0.01\\
Theoretical cross-section & 4.3 & 0.04\\
\hline
Total syst.\ unc.\ (excl.\ lumi.) & 7.1 & 0.07\\
Total syst.\ unc.\ (incl.\ lumi.) & 7.2 & 0.08\\
\hline\hline
{\bf Total uncertainty} & 7.4 & 0.08\\
\hline
 \hline
\end{tabular}
     \label{tab:vtbRes}
  \end{center}
\end{table}

\begin{figure}[htbp]
  \begin{center}
    \includegraphics[width=0.85\textwidth]{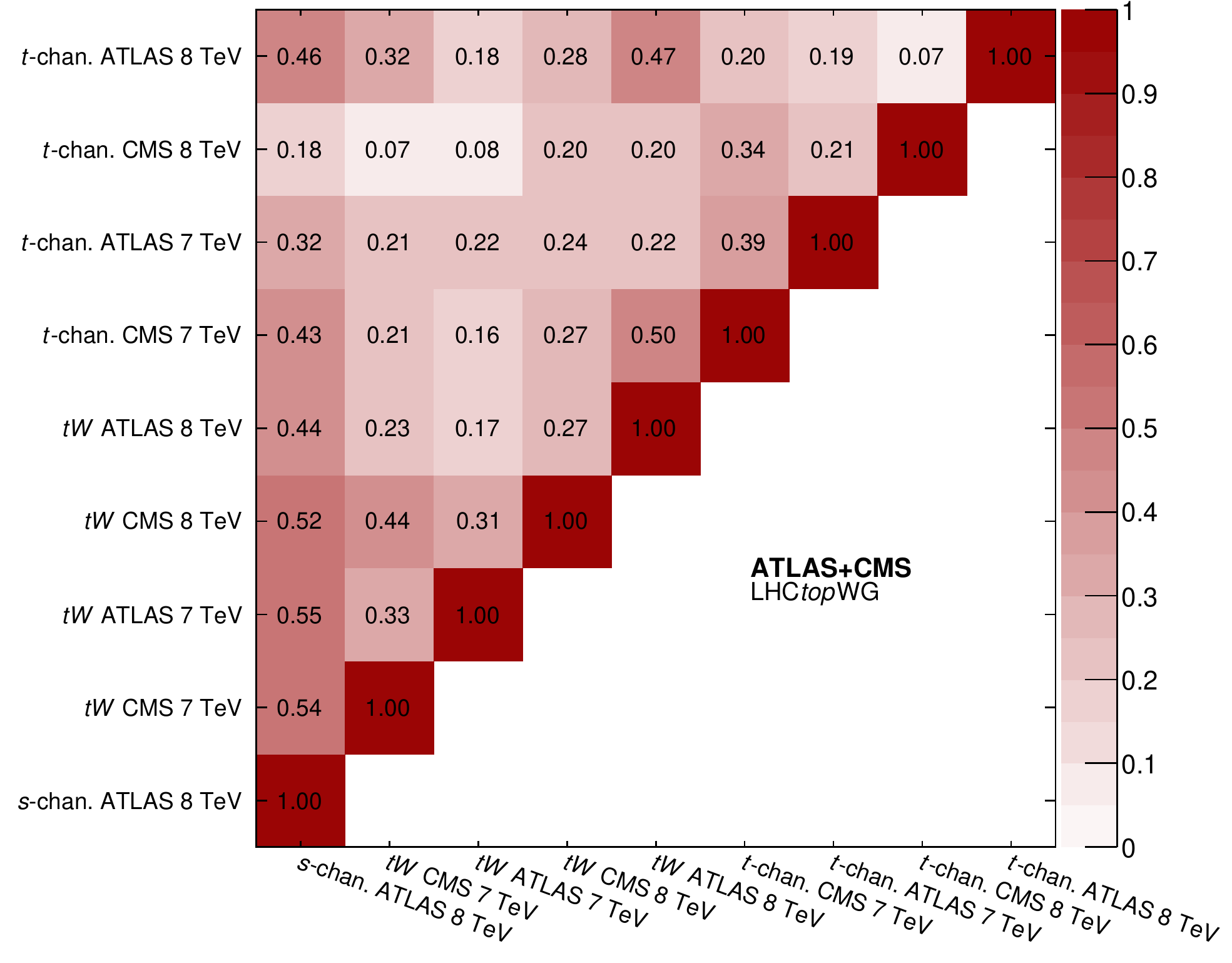}
    \caption{Correlation matrix of the overall \vtbsq\ combination. Each bin
      corresponds to a measurement in a given production mode,
      experiment, and at a given centre-of-mass energy.}
    \label{fig:rhosVtb}
  \end{center}
\end{figure}

Figure~\ref{fig:rhosVtb} illustrates the correlations between the input measurements in the combination. The correlations are all below 0.6. The largest correlations are generally between the measurements in the same experiment at the same centre-of-mass energy, and for those that have large contributions from the same theory modelling components, such as the ATLAS \sch\ measurement, which has a correlation of over 0.5 with each of the \mytW\ measurements.

The BLUE weights for each of the contributing measurements are shown in
Table~\ref{tab:wgtVtb}. The \tch\ measurements at \cmenergyEight\ have
the largest weight in the combination, followed by the \tch\ measurements at \cmenergySeven.
The $\mytW$ measurements have smaller cross-section uncertainties than the \sch\ measurements, but, in addition to the correlation between $\mytW$ and \sch\ measurements, the $\mytW$ measurements are also
more correlated with the \tch\ measurements in each experiment. The negative weights indicate the presence of large correlations between the corresponding measurement and some of the other measurements~\cite{Valassi:2013bga}.

\begin{table}[htbp]
  \begin{center}
    \caption{BLUE weights for the overall $\vtbsq$ combination.}
    \begin{tabular}{c|c|c|L{.}}
\hline
\hline
Process & $\sqrt{s}$ & Experiment & \multicolumn{1}{c}{BLUE weight} \\
\hline
\multirow{4}{*}{$t$-channel} & \multirow{2}{*}{8 TeV} & ATLAS&0.56\\
\cline{3-4}
& & CMS&0.27\\
\cline{3-4}
\cline{2-3}
 & \multirow{2}{*}{7 TeV} & ATLAS&0.07\\
\cline{3-4}
& & CMS&0.15\\
\hline
\multirow{4}{*}{$tW$} & \multirow{2}{*}{8 TeV} & ATLAS&0.05\\
\cline{3-4}
& & CMS&-0.04\\
\cline{3-4}
\cline{2-3}
 & \multirow{2}{*}{7 TeV} & ATLAS&-0.02\\
\cline{3-4}
& & CMS&0.02\\
\hline
$s$-channel & 8 TeV & ATLAS&-0.07\\
\hline\hline
\end{tabular}
     \label{tab:wgtVtb}
  \end{center}
\end{table}

The combined $\absvtb$ value from the cross-section measurements at \cmenergyComb, including uncertainties in $\sigmatheo$ for each
production mode, is
\begin{linenomath}
  \begin{equation*}
    \begin{aligned}
    \finalResultWithTheoryXSvtbComb,
    \end{aligned}
    \label{eq:result}
  \end{equation*}
\end{linenomath}
with a relative uncertainty of \relPrecisionvtbComb, which improves on the precision of 4.7\% of the most precise individual $\absvtb$ extraction, which comes from the ATLAS \tch\ analysis at \cmenergyEight~\cite{TOPQ-2015-05}. This is a 30\% improvement over the Tevatron combination~\cite{Aaltonen:2015cra}.

The $\absvtb$ values are also combined for each production mode, combining across experiments and centre-of-mass energies. For the \sch, the ATLAS and CMS measurements at \cmenergyEight\ are combined. The results are
\begin{linenomath}
  \begin{equation*}
    \begin{aligned}
      {t{\text{-channel}}}: \finalResultWithTheoryXSvtbCombtChan, \\
      {\mytW}: \finalResultWithTheoryXSvtbCombtW, \\
      {s{\text{-channel}}}: \finalResultWithTheoryXSvtbCombsChan. \\
    \end{aligned}
  \end{equation*}
\end{linenomath}
The relative uncertainties are \relPrecisionvtbCombtChan, \relPrecisionvtbCombtW\ and \relPrecisionvtbCombsChan\ respectively.  In all cases, these results
are more precise than the best individual determinations of $\absvtb$, which have uncertainties of
4.7\%, 9.9\% and 20.8\% for the \tch~\cite{TOPQ-2015-05},
\mytW~\cite{TOPQ-2012-20} and \sch~\cite{TOPQ-2015-01} analyses respectively.

\begin{figure}[!h!tbp]
  \begin{center}
    \includegraphics[width=0.82\textwidth]{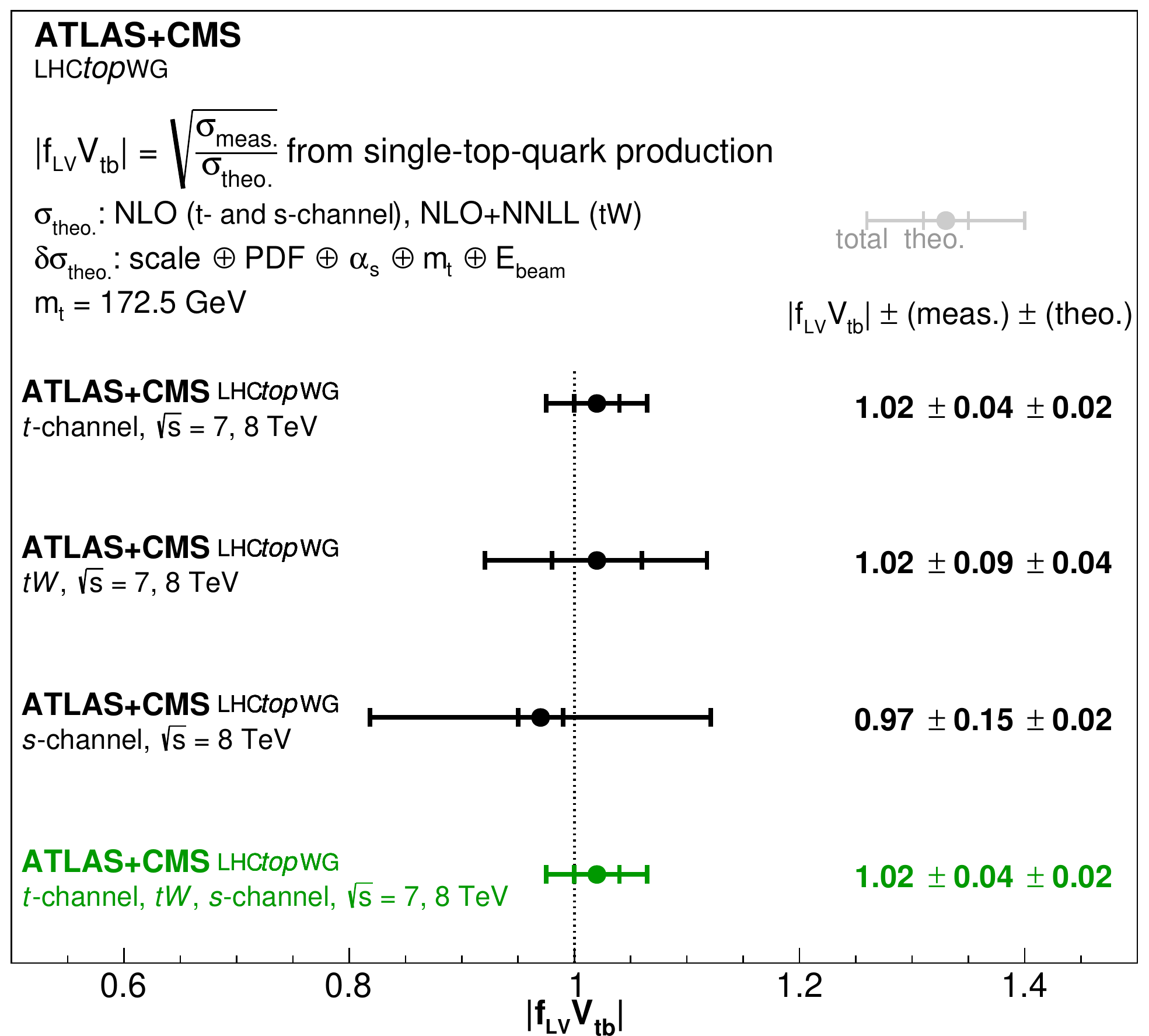}
    \caption{The combined $\absvtb$ value extracted  from the \tch\ and \mytW\ cross-section measurements at \cmenergyComb\ from ATLAS and CMS, as well as the ATLAS \sch measurement at \cmenergyEight, is shown together with the combined $\absvtb$ values for each production mode.
      The theoretical predictions for \tch\ and \sch\ production are computed at NLO accuracy, while the theoretical predictions for $\mytW$  are calculated at NLO+NNLL accuracy. The $\sigmatheo$ uncertainties used to compute $\absvtb$ include scale, PDF+$\alpha_{\rm s}$, $m_{\myt}$, and $E_{\rm beam}$ variations.
    }
    \label{fig:vtbsum}
\end{center}
\end{figure}

Figure~\ref{fig:vtbsum} shows a summary of the $\absvtb$ combinations. The combination is dominated by the \tch\ analyses.

\subsection{Stability tests}
\label{sec:tests}

\begin{figure}[!h!tbp]
  \begin{center}
    \includegraphics[width=1.0\textwidth]{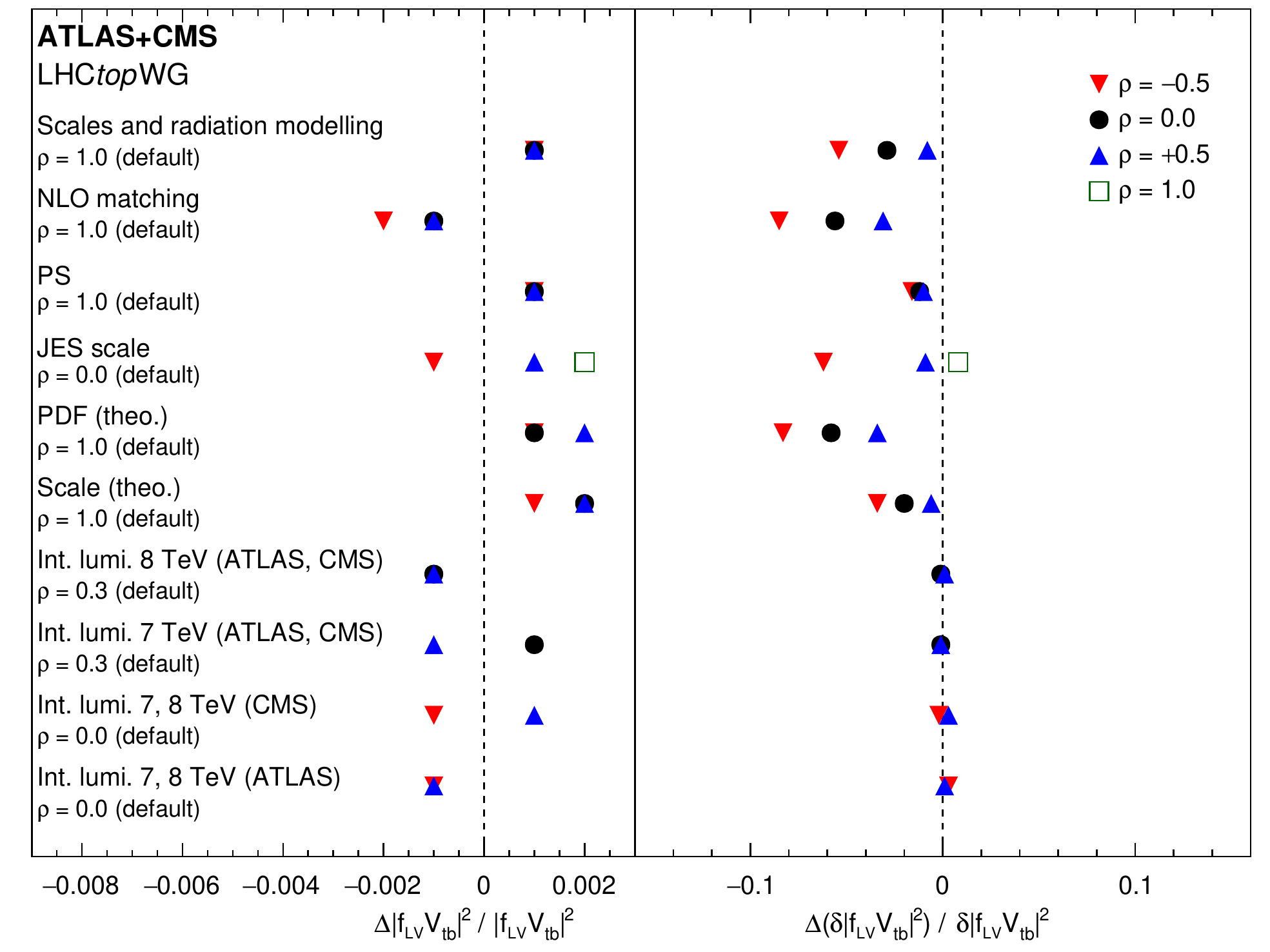}
    \caption{Results of the stability tests performed by varying
      of the correlation assumptions in different uncertainty
      categories: theory modelling (scales and radiation modelling,
      NLO matching, and PS and hadronisation), JES, dominant
      theoretical cross-section predictions (i.e.\ PDF+$\alpha_{\rm s}$
      and scale) and integrated luminosity. Two or three variations are considered depending on the
      uncertainty category. The corresponding relative shifts
      (with shift = varied $-$ nominal) in the central value, $\Delta\vtbsq/\vtbsq$, and in its uncertainty,
      $\Delta(\delta\vtbsq)/(\delta\vtbsq)$, are shown.}
    \label{fig:vtbstabilities}
  \end{center}
\end{figure}

The stability of the combination of the $\vtbsq$ values to variations in the correlation assumptions, discussed in Section~\ref{sec:systcat}, is checked for the dominant uncertainty contributions. The correlation values are varied for the theory modelling, JES, and the most important contributions to the theoretical cross-section predictions (i.e.\ PDF+$\alpha_{\rm s}$ and scale). Because of the scheme that is used for the correlations, stability tests are also performed for the uncertainties associated with the integrated luminosity. Figure~\ref{fig:vtbstabilities} summarises the results of these stability tests, where the correlations between ATLAS and CMS (and also between centre-of-mass energies for the integrated luminosity) are varied.

The uncertainties in the theory modelling category (i.e.\ scales and radiation modelling, NLO matching, and PS and
hadronisation) are varied from their default value of fully
correlated to half correlated and to the more extreme tests of uncorrelated and half anti-correlated. The JES
category is varied from its default value of uncorrelated to half correlated and half anti-correlated and the more extreme variation of fully correlated. The theoretical cross-section uncertainties, PDF+$\alpha_{\rm s}$ and
scale, are varied from their default values of fully correlated to
half correlated, uncorrelated and half anti-correlated. For the integrated luminosity, the correlation between ATLAS and CMS is varied from its default value of 30\% correlated to half correlated and uncorrelated. The correlation between different centre-of-mass energies for each experiment is varied from the default of uncorrelated to half and fully
correlated. The correlation of the theoretical scale uncertainty between different processes is also tested. For all variations, the relative changes in the central value of the combined $\absvtb$ are significantly smaller ($<$$0.5\%$) than the relative total uncertainty of \relPrecisionvtbComb. Additionally, the relative changes in the total uncertainty are below 0.004, i.e., less than $10\%$ of the total uncertainty of \totUncvtbComb. 
These tests show that the result of the combination is robust and does not critically depend on any of the correlation assumptions. The cross-section combinations similarly do not depend significantly on any of the correlation assumptions.
 
\IfPackagesLoaded{adjustbox}{\FloatBarrier}{}
\section{Summary}
\label{sec:sum}
The combinations of single-top-quark production cross-section measurements in the \tch, \myWt, and \sch\ production modes are presented, using data from LHC $pp$ collisions collected by the ATLAS and CMS Collaborations. The combinations for each production mode are performed at \cmenergyComb, using data corresponding to integrated luminosities of 1.17 to 5.1~\invfb at \cmenergySeven, and of 12.2 to 20.3~\invfb at \cmenergyEight. 
The combined \tch\ cross-sections are found to be $\finalResultTotUnctChanSeven$ and $\finalResultTotUnctChanEight$ at \cmenergyComb\ respectively. The values of the combined \myWt\ cross-sections at \cmenergyComb\ are $\finalResultTotUnctWSeven$ and $\finalResultTotUnctWEight$ respectively. For the \sch\ cross-section, the combination yields $\finalResultTotUncsChanEight$ at \cmenergyEight. 
The square of the magnitude of the CKM matrix element $\vtb$ multiplied by a form factor accounting for possible contributions from physics beyond the SM, \flv, is determined from each production mode at each centre-of-mass energy, using the ratio of the measured cross-section to its theoretical prediction, and assuming that the top-quark-related CKM matrix elements obey the relation $|\vtd|,|\vts| \ll |\vtb|$. The values of $\vtbsq$ extracted from individual ratios at \cmenergyComb\ yield a combined value of \finalResultWithTheoryXSCompactvtbComb. All combined measurements are consistent with their corresponding SM predictions.

\clearpage
\appendix
\part*{Appendix}
\addcontentsline{toc}{part}{Appendix}
\section{Systematic uncertainties in cross-section measurements}
\label{app:allUncs}
The single-top-quark cross-sections measured by the ATLAS and CMS Collaborations at \cmenergyComb, as well as the uncertainties and their correlations between experiments, are summarised in Tables~\ref{tab:tChan7XS}$-$\ref{tab:sChan8XS} for the \tch, \mytW, and \sch\ production modes. Similar to the approach that is followed in combinations using the BLUE method, the total uncertainty in these tables is evaluated as the sum in quadrature of the individual uncertainties. To obtain the impact of each source of uncertainty, the input analyses use either pseudoexperiments or approximate procedures which neglect the correlations between sources of uncertainty introduced by the fit to data. In the latter case, this may lead to small changes in the total uncertainty compared with the input measurements presented in Table~\ref{tab:xsec_meas}. The likelihood fit includes all nuisance parameters at the same time to evaluate the total uncertainty. The method used by each input analysis to evaluate the individual uncertainties is described below.

\subsection{Systematic uncertainties in \tch\ cross-section measurements}
\label{app:tch}
The \tch\ cross-sections measured by the ATLAS and CMS Collaborations at \cmenergySeven~\cite{TOPQ-2012-21,CMS-TOP-11-021} and \cmenergyEight~\cite{TOPQ-2015-05,CMS-TOP-12-038}, as well as the uncertainties and their correlations between experiments, are shown in Tables~\ref{tab:tChan7XS} and \ref{tab:tChan8XS} respectively. The total uncertainty given for each measurement is the sum in quadrature of the individual uncertainties. This is slightly different from the total uncertainty shown in Table~\ref{tab:xsec_meas} for the CMS measurements at \cmenergyComb\ since the total uncertainty is evaluated, through the fit, by varying all nuisance parameters at the same time.

In Table~\ref{tab:tChan7XS}, the CMS result at \cmenergySeven\ has
a larger data statistical uncertainty than the ATLAS result because the two experiments use data samples of different size (see Table~\ref{tab:xsec_meas}). 
In the background-normalisation category, the ``Bkg.\ from MC'' uncertainty refers to the \mytW, \sch,  \mytt, \myWZjets, and diboson backgrounds. In the ATLAS measurement, the normalisation uncertainty in the multijet background is estimated by comparing background estimates made using different techniques based on data and simulation samples, while in the CMS measurement, it is estimated from the difference between alternative methods based on data. 
There is also a large difference between the two experiments in the jets category. As discussed in Section~\ref{sec:systcat_exp}, the uncertainty in
each JES component in the ATLAS measurement is evaluated using pseudoexperiments. The CMS measurement is a BLUE combination of three different measurements, two of which introduce JES components as a nuisance parameter in the fit. Since these fits use additional control regions, the impact of the JES is reduced. In addition, the JES uncertainty in the analyses at \cmenergySeven\ is smaller for CMS~\cite{CMS-JME-10-011} than for ATLAS~\cite{PERF-2012-01}.

\begin{table}[!thp]
  \begin{center}
    \caption{Measured cross-sections, uncertainty components, their
      magnitudes (relative to the individual measurements) and the
      correlation ($\rho$) between the ATLAS and CMS $\sigmatch$
      measurements at \cmenergySeven. Uncertainties in the same row can
      be compared between experiments, as detailed in the text. The
      naming conventions follow those of the corresponding experiments.}
    \IfPackagesLoaded{adjustbox}{\begin{adjustbox}{max width=1.0\textwidth}}{}
      \begin{tabular}{l | l | r | l | r | c}
\hline
\hline
\T\B
 & \multicolumn{2}{c|}{ATLAS ($\sigmatch$, $\sqrt{s}=~$7 TeV) } &   \multicolumn{2}{c|}{CMS ($\sigmatch$, $\sqrt{s}=~$7 TeV) } & \\ 
\hline\hline
Cross-section &  \multicolumn{2}{c|}{68.0~pb} & \multicolumn{2}{c}{67.2~pb}& \\
\hline\hline
Uncertainty category &\multicolumn{2}{c|}{Uncertainty} &\multicolumn{2}{c|}{Uncertainty} &  $\rho$ \\\hline
Data statistical& &\bf 2.7\%& &\bf 5.8\% & \bf 0.0\\
\hline
\hline
Simulation statistical& &\bf 1.9\%& &\bf 1.9\% & \bf 0.0\\
\hline
\hline
Integrated luminosity& &\bf 1.8\%& &\bf 2.2\% & \bf 0.3\\
\hline
\hline
Theory modelling& Ren./fact. scales, ISR/FSR & 2.6\%& Ren./fact. scales & 3.5\% & 1.0\\
& NLO match., PS ($\mytt$, $t$-chan.) & 2.2\%& Sig. modelling (NLO method) & 4.3\% & 1.0\\
& & & Parton shower & 0.8\% & 1.0\\
& PDF & 3.2\%& PDF & 1.4\% & 1.0\\
\hline
Category subtotal & \multicolumn{2}{r|}{\bf 4.7\%} & \multicolumn{2}{r|}{\bf 5.8\%} &{\bf 0.85} \\
\hline
\hline
Background norm.& Bkg. from MC: norm. & 1.6\%& Bkg. from MC: norm. & 2.7\% & 0.0\\
& Bkg. from MC/data: multijet norm. & 1.4\%& Bkg. from data: multijet norm. & 1.3\% & 0.0\\
\hline
Category subtotal & \multicolumn{2}{r|}{\bf 2.1\%} & \multicolumn{2}{r|}{\bf 3.0\%} &{\bf 0.0} \\
\hline
\hline
Jets& JES common & 7.6\%& JES & 0.9\% & 0.0\\
& JES flavour & 1.8\%& & &0.0\\
& JetID & 1.1\%& & &0.0\\
& JER & 1.9\%& JER & 0.3\% & 0.0\\
\hline
Category subtotal & \multicolumn{2}{r|}{\bf 8.1\%} & \multicolumn{2}{r|}{\bf 0.9\%} &{\bf 0.0} \\
\hline
\hline
Detector modelling& Lepton modelling & 2.8\%& Lepton modelling & 3.5\% & 0.0\\
& & & HLT (had. part) & 1.5\% & 0.0\\
& \myMET modelling & 2.6\%& \myMET modelling & 0.1\% & 0.0\\
& \btag & 3.9\%& \btag & 2.2\% & 0.0\\
& Pile-up & 0.2\%& Pile-up & 0.6\% & 0.0\\
\hline
Category subtotal & \multicolumn{2}{r|}{\bf 5.5\%} & \multicolumn{2}{r|}{\bf 4.4\%} &{\bf 0.0} \\
\hline
\hline
Total uncertainty & \multicolumn{2}{r|}{\bf 11.7\%} & \multicolumn{2}{r|}{\bf 10.2\%} &{\bf 0.20} \\
\hline
\hline
\end{tabular}
       \IfPackagesLoaded{adjustbox}{\end{adjustbox}}{} 
    \label{tab:tChan7XS}
  \end{center}
\end{table}

\begin{table}[!ht]
  \begin{center}
    \caption{Measured cross-sections, uncertainty components, their
      magnitudes (relative to the individual measurements) and the
      correlation ($\rho$) between the ATLAS and CMS $\sigmatch$
      measurements at \cmenergyEight. Uncertainties in the same row can
      be compared between experiments, as detailed in the text. The
      naming conventions follow those of the corresponding experiments.}
    \IfPackagesLoaded{adjustbox}{\begin{adjustbox}{max width=1.0\textwidth}}{}
      \begin{tabular}{l | l | r | l | r | c}
\hline
\hline
\T\B
 & \multicolumn{2}{c|}{ATLAS ($\sigmatch$, $\sqrt{s}=~$8 TeV) } &   \multicolumn{2}{c|}{CMS ($\sigmatch$, $\sqrt{s}=~$8 TeV) } & \\ 
\hline\hline
Cross-section &  \multicolumn{2}{c|}{89.6~pb} & \multicolumn{2}{c}{83.6~pb}& \\
\hline\hline
Uncertainty category &\multicolumn{2}{c|}{Uncertainty} &\multicolumn{2}{c|}{Uncertainty} &  $\rho$ \\\hline
Data statistical& &\bf 1.4\%& &\bf 2.7\% & \bf 0.0\\
\hline
\hline
Simulation statistical& &\bf 0.8\%& &\bf 0.7\% & \bf 0.0\\
\hline
\hline
Integrated luminosity& &\bf 1.9\%& &\bf 2.6\% & \bf 0.3\\
\hline
\hline
Theory modelling& Ren./fact. scales & 3.6\%& Ren./fact. scales & 1.9\% & 1.0\\
& NLO match.  & 3.3\%& NLO match., 4FS vs 5FS & 4.9\% & 1.0\\
& Parton shower & 2.1\%& & &1.0\\
& PDF & 1.3\%& PDF & 1.9\% & 1.0\\
\hline
Category subtotal & \multicolumn{2}{r|}{\bf 5.5\%} & \multicolumn{2}{r|}{\bf 5.6\%} &{\bf 0.84} \\
\hline
\hline
Background norm.& \mytt, \mytW\ and $s$-chan. norm. & 0.1\%& $\mytt$ and \myWjets norm. & 2.2\% & 0.0\\
& Other bkg. from MC: norm. & 0.9\%& Other bkg. from MC: norm. & 0.3\% & 0.0\\
& Bkg. from MC/data: multijet norm. & 0.3\%& Bkg. from data: multijet norm. & 2.3\% & 0.0\\
\hline
Category subtotal & \multicolumn{2}{r|}{\bf 1.0\%} & \multicolumn{2}{r|}{\bf 3.2\%} &{\bf 0.0} \\
\hline
\hline
Jets& JES common & 3.2\%& JES & 4.2\% & 0.0\\
& JES flavour & 0.2\%& & &0.0\\
& JetID & 0.1\%& & &0.0\\
& JER & 0.4\%& JER & 0.7\% & 0.0\\
\hline
Category subtotal & \multicolumn{2}{r|}{\bf 3.2\%} & \multicolumn{2}{r|}{\bf 4.3\%} &{\bf 0.0} \\
\hline
\hline
Detector modelling& Lepton modelling & 1.9\%& Lepton modelling & 0.6\% & 0.0\\
& \myMET scale & 0.4\%& \myMET modelling & 0.3\% & 0.0\\
& \myMET resolution & 0.2\%& & &0.0\\
& \btag & 1.1\%& \btag & 2.5\% & 0.0\\
& Pile-up & 0.3\%& Pile-up & 0.7\% & 0.0\\
\hline
Category subtotal & \multicolumn{2}{r|}{\bf 2.3\%} & \multicolumn{2}{r|}{\bf 2.7\%} &{\bf 0.0} \\
\hline
\hline
Total uncertainty & \multicolumn{2}{r|}{\bf 7.3\%} & \multicolumn{2}{r|}{\bf 9.0\%} &{\bf 0.42} \\
\hline
\hline
\end{tabular}
       \IfPackagesLoaded{adjustbox}{\end{adjustbox}}{}  
    \label{tab:tChan8XS}
  \end{center}
\end{table}

In the analyses at \cmenergyEight, summarised in Table~\ref{tab:tChan8XS}, the difference between ATLAS and CMS in the background normalisation category is due to the different techniques used to estimate each background uncertainty. The ``Other bkg.\ from MC'' uncertainty includes the contributions from the \mytW, \sch,  \mytt, \myWZjets, and diboson backgrounds in the ATLAS analysis, and the \mytW, \sch, \myZjets, and diboson backgrounds in the CMS analysis. In the ATLAS measurement, the normalisation uncertainties associated with the top-quark, \myWZjets, diboson, and multijet backgrounds are estimated using pseudoexperiments. Variations in the theoretical cross-section predictions for these processes are also considered, except for the multijet background, where the results obtained from data and simulation samples analysed with various techniques are compared. In the CMS measurement, the uncertainty in the multijet background is estimated from the difference between alternative methods based on data. The normalisations of the \ttbar and \myWjets backgrounds are included as nuisance parameters in the fit, while the shapes of their distributions are adjusted by corrections based on data in control regions.

\FloatBarrier

\subsection{Systematic uncertainties in \mytW\ cross-section measurements}
\label{app:tw}
The \mytW\ cross-sections measured by the ATLAS and CMS Collaborations at \cmenergySeven~\cite{TOPQ-2011-17, CMS-TOP-11-022}
and \cmenergyEight~\cite{TOPQ-2012-20,CMS-TOP-12-040}, as well as the uncertainties and their correlations between experiments, are shown in Tables~\ref{tab:tW7XS} and \ref{tab:tW8XS} respectively.

In Table~\ref{tab:tW7XS}, the CMS measurement at \cmenergySeven\ takes into account the uncertainty associated with the size of the simulated event samples using the Barlow--Beeston method~\cite{Barlow:1993dm}. This contribution is included as part of the total statistical uncertainty. This uncertainty is therefore considered to be zero for the CMS measurement to avoid double-counting. Since the
statistical uncertainties in the data and simulation are uncorrelated
between the two experiments, this choice has almost no effect on the
combination. In the ATLAS analysis, the normalisation uncertainty in the misidentified lepton (fake lept.) background is conservatively taken to be 100\%, based on comparisons in data. The \myMET\ uncertainties are included in the pile-up modelling uncertainty. The \btag uncertainty is not considered because no \btag criterion is required in the event selection. The large difference in the pile-up uncertainty between ATLAS and CMS arises from different methods employed by the experiments to assess this uncertainty, as discussed in Section~\ref{sec:systcat_exp}. 

\begin{table}[!ht]
  \centering
  \caption{Measured cross-sections, uncertainty components, their
    magnitudes (relative to the individual measurements) and the
    correlation ($\rho$) between the ATLAS and CMS $\sigmatW$
    measurements at \cmenergySeven. Uncertainties in the same row can
    be compared between experiments, as detailed in the text. The
    naming conventions follow those of the corresponding experiments.}
  \IfPackagesLoaded{adjustbox}{\begin{adjustbox}{max width=1.0\textwidth}}{}
    \begin{tabular}{l | l | r | l | r | c}
\hline
\hline
\T\B
 & \multicolumn{2}{c|}{ATLAS ($\sigmatW$, $\sqrt{s}=~$7 TeV) } &   \multicolumn{2}{c|}{CMS ($\sigmatW$, $\sqrt{s}=~$7 TeV) } & \\ 
\hline\hline
Cross-section &  \multicolumn{2}{c|}{16.8~pb} & \multicolumn{2}{c}{16.0~pb}& \\
\hline\hline
Uncertainty category &\multicolumn{2}{c|}{Uncertainty} &\multicolumn{2}{c|}{Uncertainty} &  $\rho$ \\\hline
Data statistical& &\bf 17.0\%& &\bf 20.8\% & \bf 0.0\\
\hline
\hline
Simulation statistical& &\bf 2.0\%& & \bf 0.0\% &\bf 0.0\\
\hline
\hline
Integrated luminosity& &\bf 7.0\%& &\bf 4.3\% & \bf 0.3\\
\hline
\hline
Theory modelling& ISR/FSR, scales & 5.0\%& ISR/FSR, scales & 2.8\% & 1.0\\
& $\mytW$/$\mytt$ NLO match. & 10.0\%& & &1.0\\
& $\mytW$/$\mytt$ PS & 15.0\%& $\mytW$ ME/PS match. thr. & 10.1\% & 1.0\\
& PDF & 2.0\%& PDF & 2.1\% & 1.0\\
& & & DR/DS scheme & 5.9\% & 1.0\\
\hline
Category subtotal & \multicolumn{2}{r|}{\bf 18.8\%} & \multicolumn{2}{r|}{\bf 12.2\%} &{\bf 0.74} \\
\hline
\hline
Background norm.& $\mytt$ norm. & 6.0\%& $\mytt$ norm. & 6.0\% & 0.0\\
& \myZjets, diboson norm. & 8.0\%& Z/$\gamma^{*}$+jets norm. & 4.2\% & 0.0\\
& Bkg. from data: fake lept. norm. & 2.0\%& & &0.0\\
\hline
Category subtotal & \multicolumn{2}{r|}{\bf 10.2\%} & \multicolumn{2}{r|}{\bf 7.3\%} &{\bf 0.0} \\
\hline
\hline
Jets& JES & 16.0\%& JES & 15.1\% & 0.0\\
& JetID & 5.0\%& & &0.0\\
& JER & 2.0\%& JER & 3.6\% & 0.0\\
\hline
Category subtotal & \multicolumn{2}{r|}{\bf 16.9\%} & \multicolumn{2}{r|}{\bf 15.6\%} &{\bf 0.0} \\
\hline
\hline
Detector modelling& Lepton modelling & 7.0\%& Lepton modelling & 5.2\% & 0.0\\
& & & \myMET modelling & 2.5\% & 0.0\\
& & & \btag & 1.9\% & 0.0\\
& Pile-up & 10.0\%& Pile-up & 1.5\% & 0.0\\
\hline
Category subtotal & \multicolumn{2}{r|}{\bf 12.2\%} & \multicolumn{2}{r|}{\bf 6.2\%} &{\bf 0.0} \\
\hline
\hline
Total uncertainty & \multicolumn{2}{r|}{\bf 35.1\%} & \multicolumn{2}{r|}{\bf 30.6\%} &{\bf 0.17} \\
\hline
\hline
\end{tabular}
     \IfPackagesLoaded{adjustbox}{\end{adjustbox}}{}   
  \label{tab:tW7XS}
\end{table}

 In Table~\ref{tab:tW8XS}, the \mytW\ measurement by CMS at \cmenergyEight\ is based on the first half of the \cmenergyEight\ data sample. This leads
to a larger data statistical uncertainty for CMS than for ATLAS. For the same reason, the sizes of the simulated event samples are
smaller, resulting in a larger simulation statistical uncertainty in the CMS result.
In the ATLAS measurement, the normalisation uncertainty in the multijet background is estimated by comparing estimates made using different techniques on data and simulation samples, while in the CMS measurement, the uncertainty contribution of the multijet background is estimated from the difference between alternative methods based on data. In the ATLAS analysis, the misidentified lepton and non-prompt (fake lept.) background has a normalisation uncertainty of 60\%, based on comparisons in data, to account for possible mismodelling of the jet multiplicity and jet acceptance. There is a large difference between the
two experiments in the jets category. As discussed in Section~\ref{sec:systcat_exp}, the JES uncertainty in the ATLAS measurement is evaluated in different categories. The detector modelling component of the JES common uncertainty is constrained in the fit to data. In the CMS measurement, different components of the JES uncertainty are grouped together, and the group is introduced as a nuisance parameter in the fit. The \myMET\ modelling uncertainty is smaller for the CMS measurement due to the use of low-\pT jets, which allows this uncertainty to be constrained in the fit to data, as discussed in Section~\ref{sec:systcat_exp}.

\begin{table}[!ht]
  \centering
  \caption{Measured cross-sections, uncertainty components, their
    magnitudes (relative to the individual measurements) and the
    correlation ($\rho$) between the ATLAS and CMS $\sigmatW$
    measurements at \cmenergyEight. Uncertainties in the same row can
    be compared between experiments, as detailed in the text. The
    naming conventions follow those of the corresponding experiments.}
  \IfPackagesLoaded{adjustbox}{\begin{adjustbox}{max width=1.0\textwidth}}{}
    \begin{tabular}{l | l | r | l | r | c}
\hline
\hline
\T\B
 & \multicolumn{2}{c|}{ATLAS ($\sigmatW$, $\sqrt{s}=~$8 TeV) } &   \multicolumn{2}{c|}{CMS ($\sigmatW$, $\sqrt{s}=~$8 TeV) } & \\ 
\hline\hline
Cross-section &  \multicolumn{2}{c|}{23.0~pb} & \multicolumn{2}{c}{23.4~pb}& \\
\hline\hline
Uncertainty category &\multicolumn{2}{c|}{Uncertainty} &\multicolumn{2}{c|}{Uncertainty} &  $\rho$ \\\hline
Data statistical& &\bf 5.8\%& &\bf 8.1\% & \bf 0.0\\
\hline
\hline
Simulation statistical& &\bf 0.5\%& &\bf 2.4\% & \bf 0.0\\
\hline
\hline
Integrated luminosity& &\bf 4.6\%& &\bf 3.0\% & \bf 0.3\\
\hline
\hline
Theory modelling& ISR/FSR & 8.8\%& Ren./fact. scales & 12.4\% & 1.0\\
& NLO match. & 2.5\%& & &1.0\\
& Parton shower & 1.7\%& Parton shower & 14.1\% & 1.0\\
& PDF & 0.6\%& PDF & 1.7\% & 1.0\\
& $\mytW$/$\mytt$ overlap & 3.5\%& $\mytW$ DR/DS scheme & 2.1\% & 1.0\\
& & & Top-quark $p_{\mathrm{T}}$ reweight. & 0.4\% & 0.0\\
\hline
Category subtotal & \multicolumn{2}{r|}{\bf 10.0\%} & \multicolumn{2}{r|}{\bf 19.0\%} &{\bf 0.75} \\
\hline
\hline
Background norm.& $\mytt$ norm. & 1.9\%& $\mytt$ norm. & 1.7\% & 0.0\\
& \myZjets, diboson norm. & 2.0\%& \myZjets norm. & 2.6\% & 0.0\\
& Bkg. from data: fake lept. norm. & 0.3\%& & &0.0\\
\hline
Category subtotal & \multicolumn{2}{r|}{\bf 2.8\%} & \multicolumn{2}{r|}{\bf 3.1\%} &{\bf 0.0} \\
\hline
\hline
Jets& JES common & 5.3\%& JES  & 3.8\% & 0.0\\
& JES flavour & 1.9\%& & &0.0\\
& JetID & 0.2\%& & &0.0\\
& JER & 6.5\%& JER & 0.9\% & 0.0\\
\hline
Category subtotal & \multicolumn{2}{r|}{\bf 8.6\%} & \multicolumn{2}{r|}{\bf 3.9\%} &{\bf 0.0} \\
\hline
\hline
Detector modelling& Lepton modelling & 3.0\%& Lepton modelling & 1.8\% & 0.0\\
& \myMET scale & 5.5\%& \myMET modelling & 0.4\% & 0.0\\
& \myMET resolution & 0.2\%& & &0.0\\
& \btag & 1.0\%& \btag & 0.9\% & 0.0\\
& Pile-up & 2.7\%& Pile-up & 0.4\% & 0.0\\
\hline
Category subtotal & \multicolumn{2}{r|}{\bf 6.9\%} & \multicolumn{2}{r|}{\bf 2.0\%} &{\bf 0.0} \\
\hline
\hline
Total uncertainty & \multicolumn{2}{r|}{\bf 16.8\%} & \multicolumn{2}{r|}{\bf 21.7\%} &{\bf 0.40} \\
\hline
\hline
\end{tabular}
     \IfPackagesLoaded{adjustbox}{\end{adjustbox}}{}   
  \label{tab:tW8XS}
\end{table}
 
\FloatBarrier

\subsection{Systematic uncertainties in \sch\ cross-section measurements}
\label{app:sch}
The \sch\ cross-sections measured by the ATLAS and CMS Collaborations at \cmenergyEight~\cite{TOPQ-2015-01, CMS-TOP-13-009}, as well as the uncertainties and their correlations between experiments, are shown in Table~\ref{tab:sChan8XS}.

\begin{table}[!ht]
  \centering
  \caption{Measured cross-sections, uncertainty components, their
    magnitudes (relative to the individual measurements) and the
    correlation ($\rho$) between the ATLAS and CMS $\sigmasch$
    measurements at \cmenergyEight. Uncertainties in the same row can
    be compared between experiments, as detailed in the text. The
    naming conventions follow those of the corresponding experiments.}
  \IfPackagesLoaded{adjustbox}{\begin{adjustbox}{max width=1.0\textwidth}}{}
    \begin{tabular}{l | l | r | l | r | c}
\hline
\hline
\T\B
 & \multicolumn{2}{c|}{ATLAS ($\sigmasch$, $\sqrt{s}=~$8 TeV) } &   \multicolumn{2}{c|}{CMS ($\sigmasch$, $\sqrt{s}=~$8 TeV) } & \\ 
\hline\hline
Cross-section &  \multicolumn{2}{c|}{4.8~pb} & \multicolumn{2}{c}{13.4~pb}& \\
\hline\hline
Uncertainty category &\multicolumn{2}{c|}{Uncertainty} &\multicolumn{2}{c|}{Uncertainty} &  $\rho$ \\\hline
Data statistical& &\bf 16.0\%& &\bf 10.0\% & \bf 0.0\\
\hline
\hline
Simulation statistical& &\bf 12.0\%& & \bf 0.0\% &\bf 0.0\\
\hline
\hline
Integrated luminosity& &\bf 5.0\%& &\bf 4.0\% & \bf 0.3\\
\hline
\hline
Theory modelling& Ren./fact. scales & 7.0\%& Ren./fact. scales & 30.0\% & 1.0\\
& $\mytt$, $t$-chan. generator & 11.0\%& & &1.0\\
& Parton shower & 2.0\%& Parton shower & 7.0\% & 1.0\\
& PDF & 3.0\%& PDF & 7.0\% & 1.0\\
& & & Top-quark $p_{\mathrm{T}}$ reweight. & 6.0\% & 0.0\\
\hline
Category subtotal & \multicolumn{2}{r|}{\bf 13.5\%} & \multicolumn{2}{r|}{\bf 32.2\%} &{\bf 0.56} \\
\hline
\hline
Background norm.& $t$-chan., $\mytt$ norm. & 5.0\%& $t$-chan., $\mytt$ norm. & 12.0\% & 0.0\\
& \myWZjets, diboson norm. & 6.0\%& \myWZjets, diboson norm. & 12.0\% & 0.0\\
& Bkg. from data: multijet norm. & 1.0\%& Bkg. from data: multijet norm. & 2.0\% & 0.0\\
\hline
Category subtotal & \multicolumn{2}{r|}{\bf 7.9\%} & \multicolumn{2}{r|}{\bf 17.1\%} &{\bf 0.0} \\
\hline
\hline
Jets& JES common & 5.0\%& JES & 32.5\% & 0.0\\
& JES flavour & 1.0\%& & &0.0\\
& JetID & 1.0\%& & &0.0\\
& JER & 12.0\%& JER & 10.2\% & 0.0\\
\hline
Category subtotal & \multicolumn{2}{r|}{\bf 13.1\%} & \multicolumn{2}{r|}{\bf 34.1\%} &{\bf 0.0} \\
\hline
\hline
Detector modelling& Lepton modelling & 2.4\%& Lepton modelling & 1.0\% & 0.0\\
& \myMET scale & 1.0\%& \myMET modelling & 6.0\% & 0.0\\
& \myMET res & 1.0\%& & &0.0\\
& \btag & 8.0\%& \btag & 14.0\% & 0.0\\
& Pile-up & 1.0\%& Pile-up & 9.0\% & 0.0\\
\hline
Category subtotal & \multicolumn{2}{r|}{\bf 8.5\%} & \multicolumn{2}{r|}{\bf 17.7\%} &{\bf 0.0} \\
\hline
\hline
Total uncertainty & \multicolumn{2}{r|}{\bf 30.2\%} & \multicolumn{2}{r|}{\bf 54.0\%} &{\bf 0.15} \\
\hline
\hline
\end{tabular}
     \IfPackagesLoaded{adjustbox}{\end{adjustbox}}{}   
  \label{tab:sChan8XS}
\end{table}

The CMS measurement takes into account the uncertainty associated with the size of the simulated event samples using the Barlow--Beeston method~\cite{Barlow:1993dm}.  The contribution is included in the total statistical uncertainty. This uncertainty is therefore considered to be zero for the CMS measurement to avoid double-counting. Since the statistical uncertainties in the data and simulation are uncorrelated between the two experiments, this choice has almost no effect on the combination.
The result from ATLAS has smaller uncertainties. This is attributed to the use of the latest simulation samples with tuned parameters~\cite{ATL-PHYS-PUB-2015-002} as well as the use of the matrix element method in the ATLAS analysis. In addition, all systematic uncertainties are profiled in the ATLAS analysis, while in the CMS analysis, major uncertainties, including those from jets and in the theory modelling category, are excluded from the fit and evaluated using pseudoexperiments. The total uncertainties in Table~\ref{tab:sChan8XS} are slightly different from the uncertainties shown in Table~\ref{tab:xsec_meas} because here the uncertainties are summed in quadrature, while in the input analyses the impacts of at least some of the uncertainties are included in the fits to data. In particular,  the difference between the relative total uncertainty shown in Table~\ref{tab:xsec_meas} for the ATLAS measurement, i.e.\ 34.4\%, and the relative total uncertainty shown in Table~\ref{tab:sChan8XS} is due to the usage of an approximate procedure to compute the individual uncertainty contributions. Possible correlation terms between the systematic uncertainties introduced by the fit are not included here.

\section*{Acknowledgements}
We thank CERN for the very successful operation of the LHC, as well as the support staff from our institutions without whom ATLAS and CMS could not be operated efficiently.
We acknowledge the support of ANPCyT, Argentina; YerPhI, Armenia; ARC, Australia; BMWFW and FWF, Austria; ANAS, Azerbaijan; SSTC, Belarus; CNPq and FAPESP, Brazil; NSERC, NRC and CFI, Canada; CERN; CONICYT, Chile; CAS, MOST and NSFC, China; COLCIENCIAS, Colombia; MSMT CR, MPO CR and VSC CR, Czech Republic; DNRF and DNSRC, Denmark; IN2P3-CNRS, CEA-DRF/IRFU, France; SRNSFG, Georgia; BMBF, HGF, and MPG, Germany; GSRT, Greece; RGC, Hong Kong SAR, China; ISF and Benoziyo Center, Israel; INFN, Italy; MEXT and JSPS, Japan; CNRST, Morocco; NWO, Netherlands; RCN, Norway; MNiSW and NCN, Poland; FCT, Portugal; MNE/IFA, Romania; MES of Russia and NRC KI, Russian Federation; JINR; MESTD, Serbia; MSSR, Slovakia; ARRS and MIZ\v{S}, Slovenia; DST/NRF, South Africa; MINECO, Spain; SRC and Wallenberg Foundation, Sweden; SERI, SNSF and Cantons of Bern and Geneva, Switzerland; MOST, Taiwan; TAEK, Turkey; STFC, United Kingdom; DOE and NSF, United States of America. In addition, individual groups and members have received support from BCKDF, CANARIE, CRC and Compute Canada, Canada; COST, ERC, ERDF, Horizon 2020, and Marie Sk{\l}odowska-Curie Actions, European Union; Investissements d' Avenir Labex and Idex, ANR, France; DFG and AvH Foundation, Germany; Herakleitos, Thales and Aristeia programmes co-financed by EU-ESF and the Greek NSRF, Greece; BSF-NSF and GIF, Israel; CERCA Programme Generalitat de Catalunya, Spain; The Royal Society and Leverhulme Trust, United Kingdom. 
We acknowledge the enduring support for the construction and operation of the LHC and the CMS detector provided by the following funding agencies: BMBWF and FWF (Austria); FNRS and FWO (Belgium); CNPq, CAPES, FAPERJ, FAPERGS, and FAPESP (Brazil); MES (Bulgaria); CERN; CAS, MoST, and NSFC (China); COLCIENCIAS (Colombia); MSES and CSF (Croatia); RPF (Cyprus); SENESCYT (Ecuador); MoER, ERC IUT, and ERDF (Estonia); Academy of Finland, MEC, and HIP (Finland); CEA and CNRS/IN2P3 (France); BMBF, DFG, and HGF (Germany); GSRT (Greece); NKFIA (Hungary); DAE and DST (India); IPM (Iran); SFI (Ireland); INFN (Italy); MSIP and NRF (Republic of Korea); MES (Latvia); LAS (Lithuania); MOE and UM (Malaysia); BUAP, CINVESTAV, CONACYT, LNS, SEP, and UASLP-FAI (Mexico); MOS (Montenegro); MBIE (New Zealand); PAEC (Pakistan); MSHE and NSC (Poland); FCT (Portugal); JINR (Dubna); MON, RosAtom, RAS, RFBR, and NRC KI (Russia); MESTD (Serbia); SEIDI, CPAN, PCTI, and FEDER (Spain); MOSTR (Sri Lanka); Swiss Funding Agencies (Switzerland); MST (Taipei); ThEPCenter, IPST, STAR, and NSTDA (Thailand); TUBITAK and TAEK (Turkey); NASU and SFFR (Ukraine); STFC (United Kingdom); DOE and NSF (USA). \\
\hyphenation{Rachada-pisek} Individuals have received support from the Marie-Curie programme and the European Research Council and Horizon 2020 Grant, contract No. 675440 (European Union); the Leventis Foundation; the A.P.\ Sloan Foundation; the Alexander von Humboldt Foundation; the Belgian Federal Science Policy Office; the Fonds pour la Formation \`a la Recherche dans l'Industrie et dans l'Agriculture (FRIA-Belgium); the Agentschap voor Innovatie door Wetenschap en Technologie (IWT-Belgium); the F.R.S.-FNRS and FWO (Belgium) under the ``Excellence of Science -- EOS" -- be.h project n.\ 30820817; the Beijing Municipal Science \& Technology Commission, No. Z181100004218003; the Ministry of Education, Youth and Sports (MEYS) of the Czech Republic; the Lend\"ulet (``Momentum") Programme and the J\'anos Bolyai Research Scholarship of the Hungarian Academy of Sciences, the New National Excellence Program \'UNKP, the NKFIA research grants 123842, 123959, 124845, 124850, and 125105 (Hungary); the Council of Science and Industrial Research, India; the HOMING PLUS programme of the Foundation for Polish Science, cofinanced from European Union, Regional Development Fund, the Mobility Plus programme of the Ministry of Science and Higher Education, the National Science Center (Poland), contracts Harmonia 2014/14/M/ST2/00428, Opus 2014/13/B/ST2/02543, 2014/15/B/ST2/03998, and 2015/19/B/ST2/02861, Sonata-bis 2012/07/E/ST2/01406; the National Priorities Research Program by Qatar National Research Fund; the Programa Estatal de Fomento de la Investigaci{\'o}n Cient{\'i}fica y T{\'e}cnica de Excelencia Mar\'{\i}a de Maeztu, grant MDM-2015-0509 and the Programa Severo Ochoa del Principado de Asturias; the Thalis and Aristeia programmes cofinanced by EU-ESF and the Greek NSRF; the Rachadapisek Sompot Fund for Postdoctoral Fellowship, Chulalongkorn University and the Chulalongkorn Academic into Its 2nd Century Project Advancement Project (Thailand); the Welch Foundation, contract C-1845; and the Weston Havens Foundation (USA). 

In addition, we gratefully acknowledge the computing centres and personnel of the Worldwide LHC Computing Grid for delivering so effectively the computing infrastructure essential to our analyses.
In particular, the support from CERN, the ATLAS Tier-1 facilities at TRIUMF (Canada), NDGF (Denmark, Norway, Sweden), CC-IN2P3 (France), KIT/GridKA (Germany), INFN-CNAF (Italy), NL-T1 (Netherlands), PIC (Spain), ASGC (Taiwan), RAL (UK) and BNL (USA), the Tier-2 facilities worldwide and large non-WLCG resource providers is acknowledged gratefully.
Major contributors of ATLAS computing resources are listed in Ref.~\cite{ATL-GEN-PUB-2016-002}.

\printbibliography

\clearpage 
 
\begin{flushleft}
{\Large The ATLAS Collaboration}

\bigskip

M.~Aaboud$^\textrm{\scriptsize 34d}$,    
G.~Aad$^\textrm{\scriptsize 100}$,    
B.~Abbott$^\textrm{\scriptsize 126}$,    
D.C.~Abbott$^\textrm{\scriptsize 101}$,    
O.~Abdinov$^\textrm{\scriptsize 13,*}$,    
A.~Abed~Abud$^\textrm{\scriptsize 69a,69b}$,    
D.K.~Abhayasinghe$^\textrm{\scriptsize 92}$,    
S.H.~Abidi$^\textrm{\scriptsize 165}$,    
O.S.~AbouZeid$^\textrm{\scriptsize 39}$,    
N.L.~Abraham$^\textrm{\scriptsize 154}$
% ...
\bigskip
\\

% ...
$^{13}$Institute of Physics, Azerbaijan Academy of Sciences, Baku; Azerbaijan.\\
% ...
$^{34}$$^{(a)}$Facult\'e des Sciences Ain Chock, R\'eseau Universitaire de Physique des Hautes Energies - Universit\'e Hassan II, Casablanca;$^{(b)}$Centre National de l'Energie des Sciences Techniques Nucleaires (CNESTEN), Rabat;$^{(c)}$Facult\'e des Sciences Semlalia, Universit\'e Cadi Ayyad, LPHEA-Marrakech;$^{(d)}$Facult\'e des Sciences, Universit\'e Mohamed Premier and LPTPM, Oujda;$^{(e)}$Facult\'e des sciences, Universit\'e Mohammed V, Rabat; Morocco.\\
% ...
$^{39}$Niels Bohr Institute, University of Copenhagen, Copenhagen; Denmark.\\
% ...
$^{69}$$^{(a)}$INFN Sezione di Pavia;$^{(b)}$Dipartimento di Fisica, Universit\`a di Pavia, Pavia; Italy.\\
% ...
$^{92}$Department of Physics, Royal Holloway University of London, Egham; United Kingdom.\\
% ...
$^{100}$CPPM, Aix-Marseille Universit\'e, CNRS/IN2P3, Marseille; France.\\
$^{101}$Department of Physics, University of Massachusetts, Amherst MA; United States of America.\\
% ...
$^{126}$Homer L. Dodge Department of Physics and Astronomy, University of Oklahoma, Norman OK; United States of America.\\
% ...
$^{154}$Department of Physics and Astronomy, University of Sussex, Brighton; United Kingdom.\\
% ...
$^{165}$Department of Physics, University of Toronto, Toronto ON; Canada.\\
% ...
$^{a}$ Also at Borough of Manhattan Community College, City University of New York, NY; United States of America.\\
$^{b}$ Also at California State University, East Bay; United States of America.\\
$^{c}$ Also at Centre for High Performance Computing, CSIR Campus, Rosebank, Cape Town; South Africa.\\
$^{d}$ Also at CERN, Geneva; Switzerland.\\
$^{e}$ Also at CPPM, Aix-Marseille Universit\'e, CNRS/IN2P3, Marseille; France.\\
% ...
$^{*}$ Deceased

\end{flushleft}


\clearpage \vskip 5mm
\textbf{Yerevan Physics Institute, Yerevan, Armenia}\\*[0pt]
A.M.~Sirunyan, A.~Tumasyan
\vskip 5mm
\textbf{Institut f\"{u}r Hochenergiephysik, Wien, Austria}\\*[0pt]
W.~Adam, F.~Ambrogi, E.~Asilar, T.~Bergauer, J.~Brandstetter, M.~Dragicevic, J.~Er\"{o}, A.~Escalante~Del~Valle, M.~Flechl, R.~Fr\"{u}hwirth$^{1}$, V.M.~Ghete, J.~Hrubec, M.~Jeitler$^{1}$, N.~Krammer, I.~Kr\"{a}tschmer, D.~Liko, T.~Madlener, I.~Mikulec, N.~Rad, H.~Rohringer, J.~Schieck$^{1}$, R.~Sch\"{o}fbeck, M.~Spanring, D.~Spitzbart, W.~Waltenberger, J.~Wittmann, C.-E.~Wulz$^{1}$, M.~Zarucki
% ...
\vskip 5mm
% ...
1:  Also at Vienna University of Technology, Vienna, Austria\\
% ...
13: Also at Zewail City of Science and Technology, Zewail, Egypt\\
% ...
34: Also at Kyung Hee University, Department of Physics, Seoul, Korea\\
% ...
39: Also at Warsaw University of Technology, Institute of Electronic Systems, Warsaw, Poland\\
% ...
69: Also at Monash University, Faculty of Science, Clayton, Australia\\
% ...

\end{document}